\documentclass[aps,prd,twocolumn,showpacs,superscriptaddress,nofootinbib]{revtex4-1}
\usepackage{graphicx}  %
\usepackage{dcolumn}   %
\usepackage{bm}        %
\usepackage{amssymb}   %
\usepackage{color}
\usepackage{aas_macros}
\usepackage{xspace}
\usepackage[normalem]{ulem}

\usepackage{subcaption}
\usepackage{amsmath}
\usepackage{caption}
\usepackage{multirow}
\newcommand{\msun}{\ensuremath{\mathrm{M}_\odot}}

\newcommand{\Event}{GW150914}
\newcommand{\Xmas}{GW151226}
\newcommand{\Second}{LVT151012}
\newcommand{\BNS}{GW170817}

\newcommand{\WF}{{\tt IMRPhenomPv2}}

\newcommand{\si}{\ensuremath{\sim}\xspace}

\newcommand{\chip}{\ensuremath{\chi_\mathrm{p}}\xspace}
\newcommand{\chieff}{\ensuremath{\chi_\mathrm{eff}}\xspace}

\newcommand{\beq}{\begin{equation}}
\newcommand{\eeq}{\end{equation}}
\newcommand{\degg}{deg\ensuremath{^2}\xspace}

\newcommand{\TGLE}{\ensuremath{\mathrm{L}_\mathrm{CE} \mathrm{E}_1}\xspace}
\newcommand{\TGE}{\ensuremath{\mathrm{E}_1}\xspace}
\newcommand{\TGEE}{\ensuremath{\mathrm{E}_1 \mathrm{E}_2}\xspace}
\newcommand{\TGL}{\ensuremath{\mathrm{L}_\mathrm{CE}}\xspace}
\newcommand{\LE}{\ensuremath{\mathrm{L}_\mathrm{V} \mathrm{E}_1}\xspace}
\newcommand{\LEI}{\ensuremath{\mathrm{L}_\mathrm{V}\mathrm{I}_\mathrm{V} \mathrm{E}_1}\xspace}

\begin{document}

\title{Characterization of binary black holes by heterogeneous gravitational-wave networks}
\author{Salvatore Vitale}
\email{salvatore.vitale@ligo.org}
\author{Chris Whittle}

\affiliation{LIGO, Massachusetts Institute of Technology, Cambridge, Massachusetts 02139, USA}

\date{\today}
\begin{abstract}

Gravitational waves detected by advanced ground-based detectors have allowed studying the universe in a way which is fully complementary to electromagnetic observations.
As more sources are detected, it will be possible to measure properties of the local population of black holes and neutron stars, including their mass and spin distributions.
Once at design sensitivity, existing instruments will be able to detect heavy binary black holes at redshifts of \si 1. 
Significant upgrades in the current facilities could increase the sensitivity by another factor of few, further extending reach and signal-to-noise ratio.
More is required to access the most remote corners of the universe. Third-generation gravitational-wave detectors have been proposed, which could observe most of the binary black holes merging anywhere in the universe. 
In this paper we check if and to which extent it makes sense to keep previous-generation detectors up and running once a significantly more sensitive detector is online.
First, we focus on a population of binary black holes with redshifts distributed uniformly in comoving volume. We show that measurement of extrinsic parameters, such as sky position, inclination and  luminosity distance can significantly benefit from the presence of a less sensitive detector. Conversely, intrinsic parameters such as \emph{detector-frame} masses and spins are largely unaffected. Measurement of the \emph{source-frame masses} is instead improved, owing to the improvement of the distance measurement.
Then, we focus on nearby events. We simulated sources similar to GW150914 and GW151226 and check how well their parameters can be measured by various networks. Here too we find that the main difference is a better estimation of the sky position, although even a single triangular-shaped third-generation detector can estimate their sky position to 1~\degg or better. 
\end{abstract}

\maketitle

\section{Introduction}
The detection of 5 binary black hole (BBH) mergers by the Advanced LIGO~\cite{Harry:2010zz} and Virgo~\cite{2015CQGra..32b4001A} observatories, has started the field of gravitational-wave astrophysics~\cite{GW150914-DETECTION,GW151226-DETECTION,2017ApJ...851L..35A,2017PhRvL.119n1101A,2017PhRvL.118v1101A}.
The local merger rates of BBH inferred from the first two observing runs, $12-213$ Gpc$^{-3}$ yr$^{-1}$~\cite{2017PhRvL.118v1101A}, imply tens to hundreds of BBHs will be detected in the next few years as Advanced LIGO progresses toward its design sensitivity~\cite{2016LRR....19....1A}. 
Detecting large numbers of black holes in binary systems will shed light on the formation channels of compact binaries~\cite{Vitale:2015tea,2017MNRAS.471.2801S,2017PhRvD..96b3012T,2017Natur.548..426F,2017arXiv170907896F}, the role of natal kicks~\cite{2000ApJ...541..319K,2008ApJ...682..474B,2013PhRvD..87j4028G,2016ApJ...832L...2R}, and on the mass function of stellar mass black holes, including the existence and characteristics of a mass gap~\cite{2012ApJ...757...36K,2016arXiv160808223M}.

As the network of gravitational-wave (GW) observatories grows in the next few years, we will see improvements in measuring sources' sky position~\cite{2011CQGra..28j5021F,2011PhRvD..84j4020V,2016LRR....19....1A,VeitchMandel:2012,2014arXiv1404.5623S} and polarizations.
This has already been demonstrated by Advanced Virgo. The two GW sources detected by the three-detector network of the two LIGOs and Virgo have been localized with uncertainties of 60~\degg~\cite{2017PhRvL.119n1101A} and 28~\degg~\cite{2017PhRvL.119p1101A}, as opposed to the hundreds or thousands of square degrees achievable with only two detectors.
The first three-detector discovery, the BBH GW170814, allowed for tests of the polarization content of gravitational-wave signals~\cite{2017PhRvL.119n1101A}.
The improved sky localization of the binary neutron star (BNS) GW170817~\cite{2017PhRvL.119p1101A} spectacularly paid off with the identification of the host galaxy and the  discovery of electromagnetic counterparts at all frequencies~\cite{2017ApJ...848L..12A}.

The construction phase of the Japanese detector KAGRA~\cite{2012CQGra..29l4007S,PhysRevD.88.043007} has finished, and the instrument could join the global network in the early 2020s. Another LIGO instrument, LIGO India~\cite{M1100296,Indigo} will be built, and could be online in the mid 2020s.

These advanced detectors will be sensitive to BBH up to a redshift of \si1 (the exact limit depending on the specific mass function one considers, and in particular on whether intermediate-mass black holes exist). Straightforward updates such as the implementation of squeezed states of light~\cite{LIG11b,Bar13a} can increase the strain sensitivity of advanced detectors by a factor of \si2~\cite{2015PhRvD..91f2005M} (this enhanced configuration is often referred to as ``Advanced LIGO+'', or simply A+).

LIGO Voyager (henceforth just Voyager) is a proposed evolution of the LIGO detectors in which silicon test mass are used, kept at a moderate level of cryogeny~\cite{Voyager}.%
Voyager could be implemented in the current LIGO facilities, and would have a strain sensitivity a factor of \si 2 better than A+ (see Fig.~\ref{Fig.PSD} below). The possibility of re-using most of the advanced detectors infrastructure results in a moderate incremental cost for building Voyager-class instruments to replace advanced ones.

At the same time, intense R\&D is ongoing to verify what sensitivity can be achieved building new infrastructure (third-generation, or 3G detectors). Two main designs are currently under consideration.

The Einstein Telescope (ET) observatory~\cite{2010CQGra..27s4002P} is comprised of three 10 km long interferometers with inter-arm angles of 60$^\circ$, arranged as a triangle.
The departure from the L-shaped geometry used by gravitational-wave detectors to date allows for some polarization resolution, even with only a single 3G detector online~\cite{2009CQGra..26h5012F}.
In contrast, a single L-shaped detector would be unable to distinguish between gravitational-wave polarizations.
The ET detector is planned to be constructed underground, thereby retaining good sensitivity all the way down to a couple of Hz, as opposed to the \si10 Hz typically achievable by ground-based detectors. 

The Cosmic Explorer (CE) observatory~\cite{2016arXiv160708697A,PhysRevD.91.082001} keeps the traditional shape and extends the arm length to 40 km in order to improve sensitivity by directly increasing the displacement caused by gravitational waves.

3G detectors would have a strain sensitivity roughly a factor of 10 better than advanced detectors, hence greatly increasing the observable fraction of universe. 

The science goals achievable with 3G detectors have been extensively discussed elsewhere, with a focus on the science possible with only the ET observatories due to the longer history of analysis in comparison with the CE detector, a more recently conceptualized detector.
For example, Ref.~\cite{2010CQGra..27u5006S,2011PhRvD..83b3005Z,2012PhRvD..86b3502T,2016PhRvL.117j1102B} reported on studies of cosmology, cosmography and black holes spectroscopy.
Binary neutron star detection and characterization have been explored by Ref.~\cite{PhysRevD.86.122001,2016PhRvD..93b4018M}, while Ref.~\cite{Mishra:2010tp,2014PhRvD..90f4009M,2012PhRvD..85l4056G,2016PhRvD..94h4024B,2017arXiv170408268C} focus on tests of general relativity.

The capabilities of 3G networks to estimate parameters of BBH has been explored in Ref.~\cite{Vitale3G}, showing that 3G detectors will be able to detect BBH all the way to redshift of \si10 and above.
The orbit of these sources will be isotropically oriented~\cite{2016PhRvD..94l1501V}, which will make eventual spin-induced precession visible in the detector frame.
Additionally, the fact that a significant fraction of events will be loud implies that the parameters of sources can be estimated extremely well all the way to redshift of a few.
A striking difference from existing detectors will be the quality of spin estimation, currently quite poor~\cite{PhysRevLett.112.251101,Vitale:2016avz}.
Finally, Ref.~\cite{2017PhRvL.118o1105R} has shown how networks of 3G detectors would easily facilitate the detection of a stochastic signal comprised of all the individually-unresolvable sources.

The minimum or the optimal number of 3G detectors needed does of course depend on the specific science goal being pursued. This question has been addressed by Ref.~\cite{Vitale3G} in the context of BBH characterization.

In this paper we explore a different question: once a 3G detector is operational, would previous-generation Voyager instruments be useful, or would they instead not significantly improve upon the science that the more sensitive 3G interferometer can deliver?

Ref.~\cite{2017arXiv170800806M} deals with this question in the context of localizing binary neutron star mergers using timing, amplitude and phase information.

In contrast, we cast our attention to BBH and fully evaluate the posterior distributions of the unknown source parameters using Bayesian parameter estimation~\cite{2015PhRvD..91d2003V}. To this end, we simulate a population of BBH with positions uniformly distributed in comoving volume.

To contain the number of simulations to be performed, we focus on a specific scenario: one in which the ET is the first 3G detector to go online.
While there is no guarantee that will be the case, it is a reasonable assumption since the design of ET has been developed for much longer than the CE's.

We will thus compare the performances of a single ET-type site with that of an ET and one or two Voyager-class instruments. This will give us a quantitative assessment of whether Voyager can add anything to a single 3G instrument.  
We will also consider networks of two 3G instruments (one CE and one ET, and two ETs) to give an idea of how heterogenous networks would compare with a network of two 3Gs (networks of two CEs, or more than two 3G sites have already been considered by Ref.~\cite{Vitale3G}).

We find that adding even a single Voyager improves the uncertainty in the estimation of the source sky positions by roughly a factor of 100 (compared to a single ET), although we can gain a further factor of \si10 with the addition of another 3G detector. We also observe significant improvement for the estimation of the orbital inclination angle.
Smaller improvements are visible for the estimation of the luminosity distance and the source-frame mass parameters.
The estimation of spins is not significantly improved by the addition one or two Voyagers, but does get better if a second 3G detector is added.

The population of astrophysically distributed events is such that most sources are located at redshifts of order unity, which resulted in no simulated events in our population at redshifts of \si $0.1$, the distance at which the advanced instruments made the first discoveries. We thus simulate software copies of GW150914~\cite{GW150914-DETECTION} and GW151226~\cite{GW151226-DETECTION} and analyze them under the various networks we described above. We find that even a single ET can localize these nearby loud sources within \si~1~\degg. 
We show that 3G detectors can estimate the masses of a GW150914-like (GW151226-like) event with 90\% uncertainty \si $1000 \,(100)$ smaller than what LIGO measured~\cite{O1-BBH}.
For the mass ratio and the effective spin parameter, the improvement is of the order of \si$100 (10)$.
For the distance and sky location, the number of detectors in the network matters more. While a single CE cannot localize events well, a single ET can deliver measurements of distance and sky position which are a factor of \si 10 and 100, respectively, better than what was obtained with advanced detectors.
We find that introducing Voyager detectors offers negligible improvement in intrinsic parameters of the sources, while sky localization performs better by up to two orders of magnitude.

The rest of this paper is organized as follows.
Section~\ref{sec:networks} describes the 3G and heterogeneous networks used in this study.
In Section~\ref{sec:results}, we explore the results for certain key parameters of BBHs in a population of events.
In Section~\ref{sec:nearby}, we investigate how the addition of Voyager(s) or 3G detectors to a network improve the parameter estimation for particularly loud BBHs, so-called ``gold-plated events" such as GW150914.
Concluding remarks are included in Section~\ref{sec:conclusions}.
 
\section{Networks and simulated sources}\label{sec:networks}

We consider two representative heterogeneous networks. 

\begin{itemize}
\item \LE: An Einstein Telescope at the Virgo site (the final site of ET is still to be finalized, but the results won't change much as a function of its location in Europe), and one Voyager detector at the LIGO Livingston site.
\item \LEI: An Einstein Telescope at the Virgo site, one Voyager detector at the LIGO Livingston site and one Voyager in India.
\end{itemize}

We will compare their performance against three 3G networks.

\begin{itemize}
\item \TGE: A single Einstein Telescope at the same position of the Virgo detector.
\item \TGLE: An Einstein Telescope at the Virgo site and one Cosmic Explorer at the LIGO Livingston site.
\item \TGEE: A network with two Einstein Telescopes, one at the same position as Virgo and the other at the Hanford site.
\end{itemize}

\begin{figure}[htb!]
\includegraphics[width=0.48\textwidth]{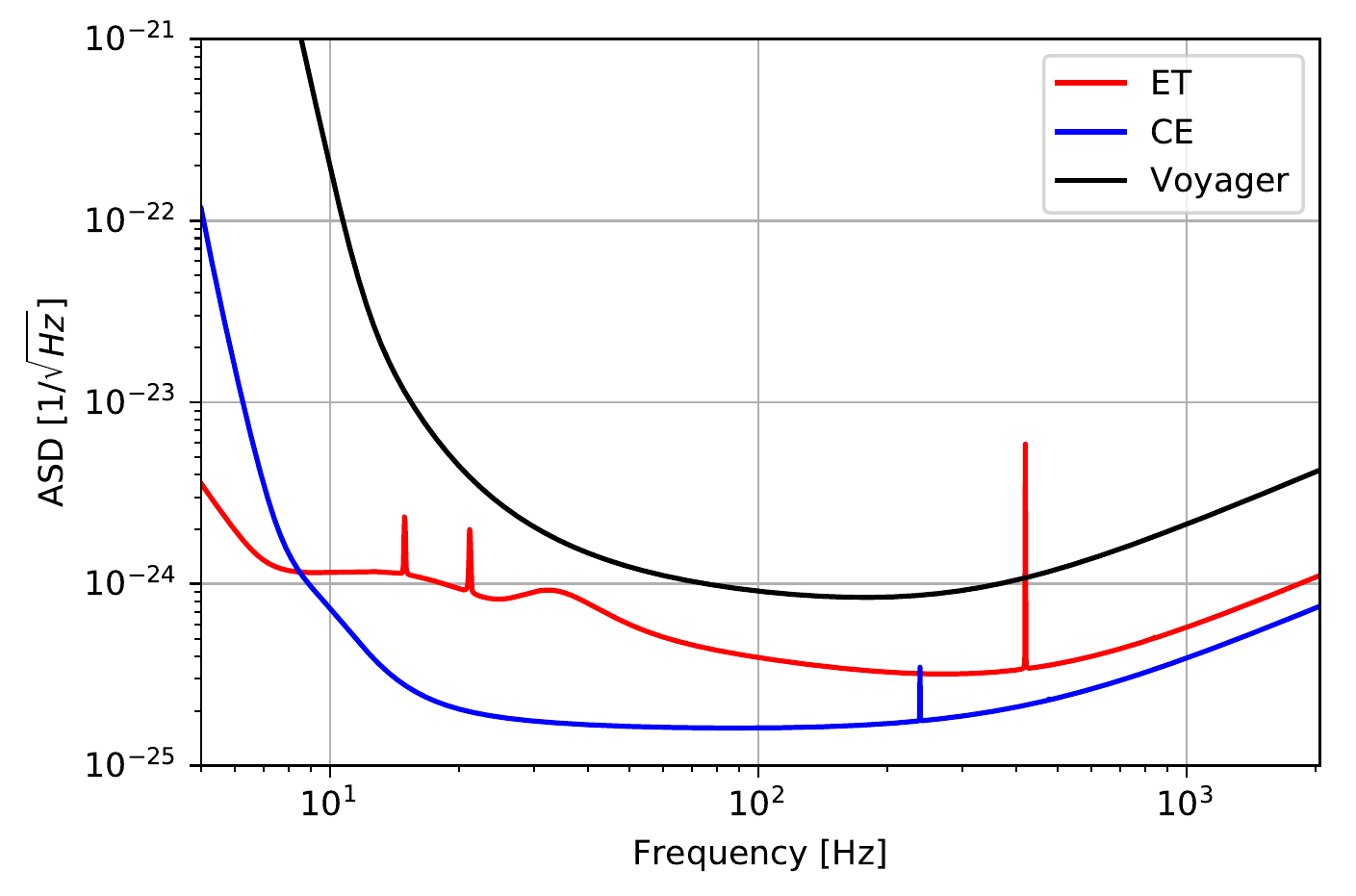}
\caption{Amplitude spectral density of the design sensitivity for Voyager and the 3G detectors considered here.}
\label{Fig.PSD}
\end{figure}

In Fig.~\ref{Fig.PSD} we report the projected noise amplitude spectral densities (ASD) for each class of instrument. For ET we used the ET-D design~\cite{2010CQGra..27s4002P}; for CE we used the standard 40-Km configuration~\cite{2016arXiv160708697A}; for Voyager the noise density reported in Ref.~\cite{Voyager}.

In this work we do not consider the proposed A+ update to LIGO~\cite{2015PhRvD..91f2005M}. However, since the improvement in strain sensitivity of LIGO vs. A+ is similar to the improvement of Voyager vs. ET, one can expect that the main conclusions we obtain in this paper would be applicable to heterogeneous networks of Advanced LIGO and A+.

Although the ET-D design would have sensitivity down to the Hz region, to contain the computational cost we started the analysis from $10$~Hz for all networks. Since ET features in all networks,  and we are mostly interested in how networks compare, this limitation is not likely to play a significant role.

For all networks, we generated a population of BBH following the same procedure described in Ref.~\cite{Vitale3G}, which we quickly summarize here.

\begin{figure}[htb]
\begin{tabular}[c]{cc}
\centering
\begin{subfigure}{0.45\textwidth}
\includegraphics[width=\textwidth]{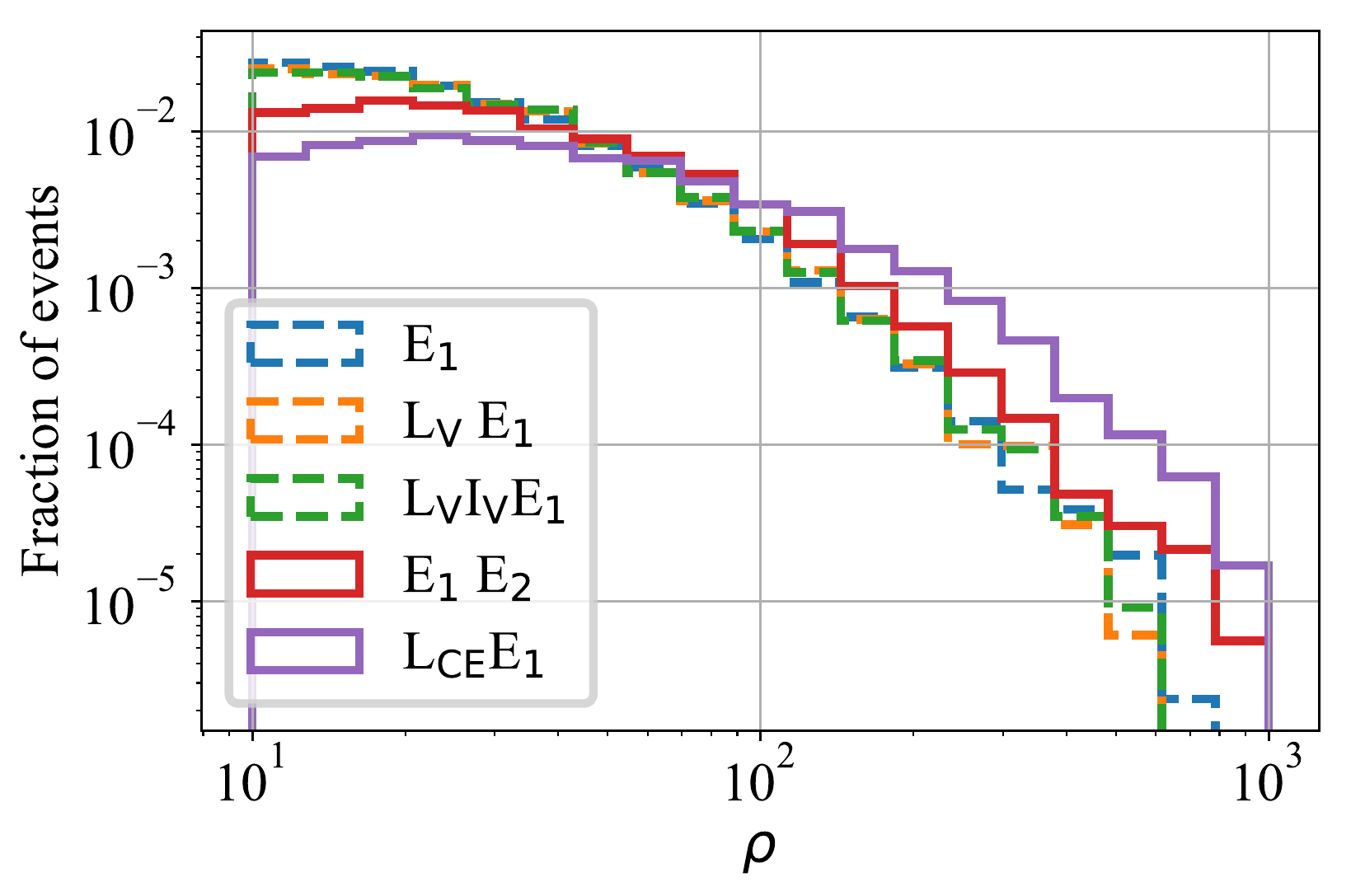}
\end{subfigure} &\\
\begin{subfigure}{0.45\textwidth}
\includegraphics[width=\textwidth]{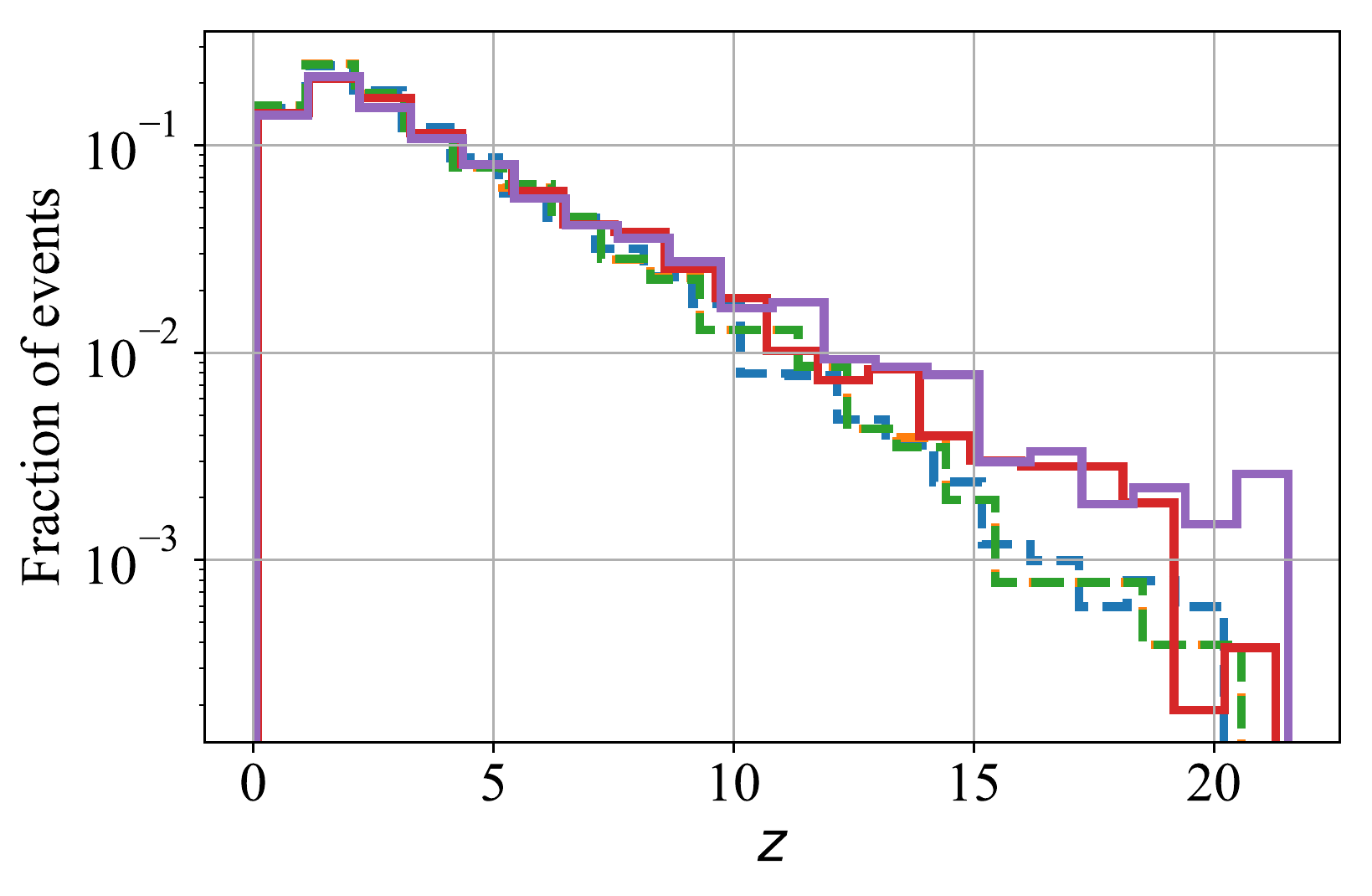}
\end{subfigure} 
\end{tabular}\caption{Distribution of network SNRs $\rho$ (\emph{above}) and redshifts $z$ (\emph{below}) for the networks considered in this work. Networks with a single 3G instruments are indicated with dashed lines.}\label{fig:redshifts}
\end{figure}

A random set of intrinsic and extrinsic parameters is drawn from the relevant distributions.
The \emph{source-frame} total mass is uniform in the range $[12-200]$~\msun. The magnitude of the dimensionless spin is random in the range $[0,1]$ (where 0 means no spin, and 1 is the maximal spin).
The spin direction, sky position and orbital orientation are all uniform in the unit sphere. Polarization and coalescence phase are uniform in the appropriate range. 
Finally, a redshift is randomly drawn uniformly in comoving volume, in the range $[0,22]$, using a standard $\Lambda$CDM cosmology~\cite{2015arXiv150201589P}. 
After all parameters are drawn, the corresponding GW signal is generated and added into simulated noise from each detector and its network signal-to-noise ratio (SNR)~\cite{SathyaSchutzLRR} is calculated. If the SNR is above a reasonable detection threshold (10) the parameters are stored; if not, the whole procedure is repeated.
For each network, 5,000 systems are generated this way. In Fig.~\ref{fig:redshifts} we report the distribution of  network SNR (top) and redshift (bottom) for all catalogs.

We see that \TGE, \LE and \LEI basically return the same distributions, compatible with the fact that for most events the Voyager detectors would contribute relatively little to the SNR.
On the other hand, the \TGEE and \TGLE networks can clearly detect sources farther away, and at higher SNRs.
For all networks, however, we notice that most sources will come from redshifts of \si$[1-2]$ where more volume is available~\cite{1999astro.ph..5116H}.
These will constitute the bulk of the population of BBH detected by future instruments. 
 
 From these catalogs, \si 210 (randomly generated) sources are selected for each network, and their parameters are estimated using a stochastic sampling~\cite{2015PhRvD..91d2003V}.
 The main outcome of this algorithm is a posterior distribution of the unknown parameters on which each event depends, $\vec{\theta}$, given the data from all the instruments in the network.
We used the  \WF{} waveform model~\cite{2015PhRvD..91b4043S,Hannam:2013oca}, both for simulating the waveforms added into synthetic noise and as templates while calculating the likelihood~\cite{2015PhRvD..91d2003V}. 

We remind that \WF{} does not contain higher-order amplitude corrections to the waveform. If has been shown that inclusion of higher order mode can improve the measurement of some parameters, especially for systems with high mass ratios and large inclinations~\cite{2017PhRvD..95j4038C,2017PhRvD..96l4024V,2016PhRvD..93h4019C,2015PhRvD..92b2002G,2017arXiv170800404L}. Our results can thus be considered conservative estimates.

\section{Results}\label{sec:results}

In this section, we summarize the results of the astrophysical BBH simulations, and how well sources can be characterized by each network.
Given that the strain sensitivity of Voyager-class detectors is a factor of few lower than ET's (Fig.~\ref{Fig.PSD}), we expect that heterogeneous networks won't yield significantly better estimation for parameters whose precision depends on the SNR more than on the size and geometry of the network~\cite{Vitale3G}.

These are parameters such as detector-frame masses and spins, which can be measured well even with a single instrument. Conversely, \emph{extrinsic} parameters such as distance, sky position and inclination, do require information about polarization and/or timing to be measured. Thus, those will benefit the most from having more than one site.

Unless otherwise specified, we report  90\% confidence intervals; occasionally, the 90\% confidence interval will be relative to the true value, and measured in \%.

\subsection{Distance and sky position}

Gravitational-wave detectors are sensitive to all directions (although not exactly with the same sensitivity). Information about a source's position is thus mostly determined by time-triangulation~\cite{Fairhurst2009}. Additionally, the relative amplitudes of the gravitational-wave signals in the various detectors of a network also provide information about the sky location.
Adding a second or third detector can thus dramatically reduce the uncertainty in the localization of the sources.

It is important to notice how non-detections in one or more instruments of a network \emph{can} help sky localization. This happens because detectors are not sensitive to all directions in the same way. Instead, L-shaped detector's antenna patterns have a characteristic ``peanut" shape~\cite{2011CQGra..28l5023S}, which clearly shows how detectors are more sensitive to sources directly overhead.
A non-detection in one detector can thus still be useful by excluding regions of the sky where that detector's antenna pattern is large and could potentially have led to a detectable signal.

This was the case for the BNS detection \BNS: the low-latency localization improved from 190 deg$^2$ to 31 deg$^2$ when Virgo was included in the analysis, despite an SNR of only 2.0 in the Virgo data~\cite{2017PhRvL.119p1101A}.

\begin{figure*}[htb!]
\includegraphics[width=\textwidth]{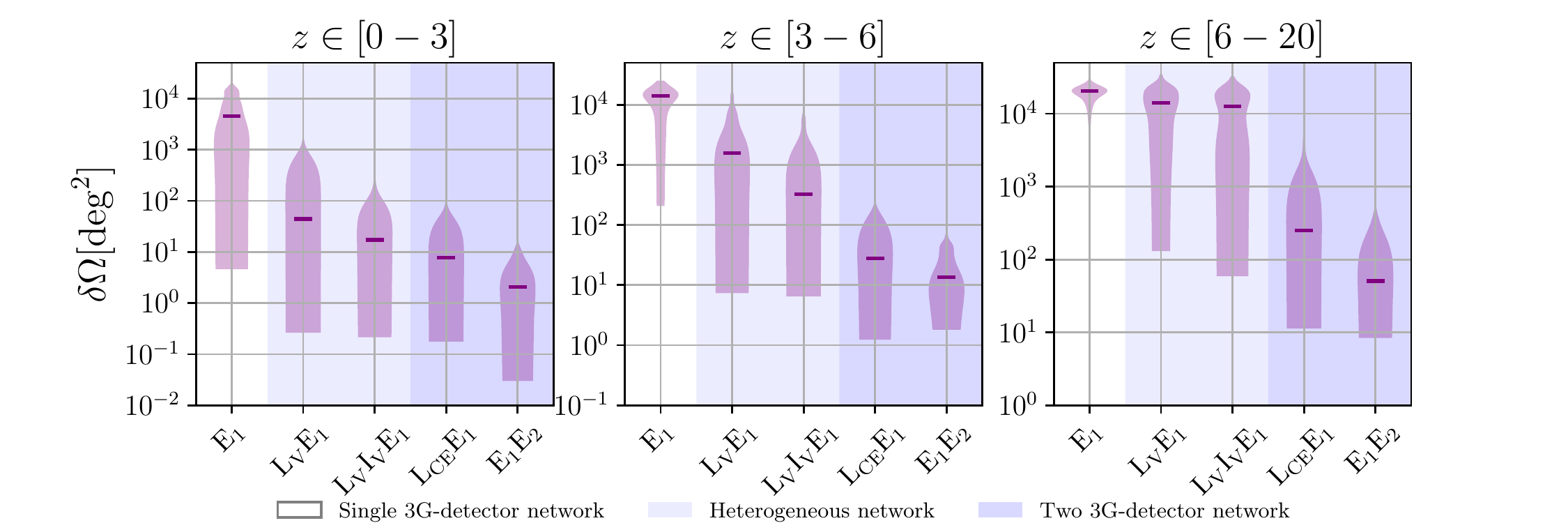}
\caption{Violin plots for the 90\% uncertainties for the sky localization (\degg, y axis) obtained from each network (x axis).}
\label{fig:sky}
\end{figure*}

Fig.~\ref{fig:sky} shows violin plots for the 90\% credible region of the sky positions of the BBH population.
We see that a network comprised of a single ET is unable to localize all but the very nearest and loudest events.
For such nearby events, accurate localization leans heavily on the ability of the ET detector to extract polarization information.
The typical BBH detection with an \TGE detector would have a sky uncertainty of $\gtrsim 500$~\degg. 
This is comparable to the Advanced LIGO's  localization of GW150914: 600 deg$^2$~\cite{GW150914-DETECTION}.
Any localizing ability of the one-detector \TGE network almost entirely disappears as we move to redshifts of $z > 3$.

Including a single LIGO Voyager detector alongside the ET observatory improves the mean localization by approximately two orders of magnitude for nearby ($z<3$) events.
The \LE network can localize nearby events to areas of tens of \degg on average, with the best events localized within areas of $10$~\degg and better. 
The second Voyager detector further reduces the number of poorly localized events, and brings the median uncertainty down to $\sim 20$~\degg.

Similar trends can be seen for sources at higher redshifts, although the improvement offered by heterogeneous networks is smaller. In the $z\in [3,6]$ bin, the sky uncertainty goes down by one order of magnitude for \LE compared to \TGE, and a factor of \si50 for \LEI compared to \TGE. The fact that the improvement is more limited for remote events can be trivially explained with the fact that a large fraction of BBH at these distances would be too weak for a non-detection in Voyager to be useful: even if placed overhead Voyager, they would still be undetectable.
For the furthest BBHs, the Voyager(s) enable only localization for the loudest few events, with minimal localizing ability for the rest of the population.

At all redshifts, multi-detector 3G networks offer additional improvement, with the \TGEE network consistently outperforming \TGLE.
We see that the extra polarization information provided by a second ET detector results in a greater improvement than the CE observatory, despite the latter instrument typically yielding a higher SNR.
This is yet another way of saying that geometry matters more than SNR when it comes to localizing GW sources.

With \TGEE and \TGLE networks, one can localize most of the BBH population up to high redshifts.

\begin{figure}[htb!]
\includegraphics[width=0.49\textwidth]{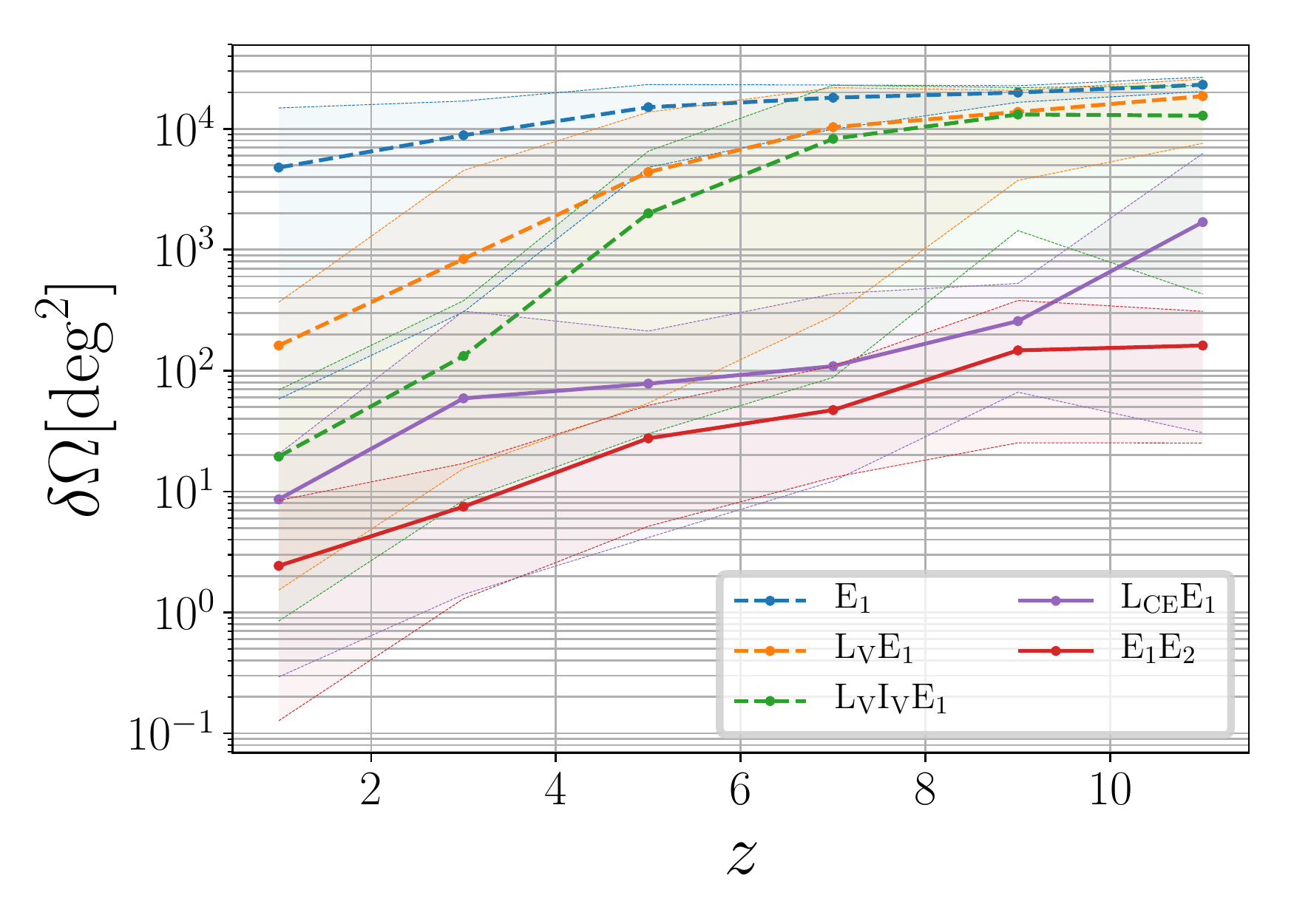}
\caption{90\% uncertainties for the sky localization (\degg, y axis) a function of redshift (x axis). The thicker lines report the mean uncertainty in each redshift bin, while the dotted lines denote the region in which confidence intervals fall for 90\% of events in that redshift bin.}
\label{fig:distance_med}
\end{figure}

In Fig.~\ref{fig:distance_med} we provide a different representation of the 90\% uncertainty in the sky location for the various networks as a function of the redshift.

\begin{figure*}[htb!]
\includegraphics[width=\textwidth]{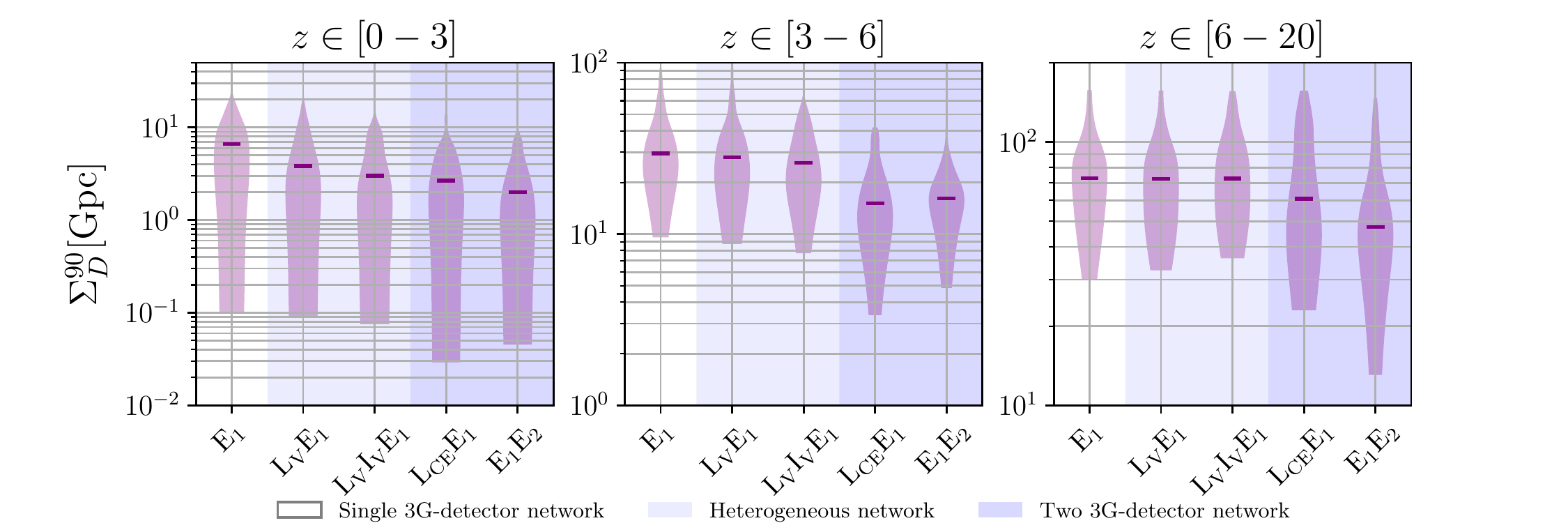}
\caption{Violin plots for the 90\% relative uncertainties for the luminosity distance (y axis) obtained from each 3G network (x axis).}
\label{fig:distance}
\end{figure*}

Estimation of the BBH luminosity distance sees a much less dramatic improvement with the addition of Voyagers or a second 3G detector.
Fig.~\ref{fig:distance} shows violin plots for the 90\% confidence interval for the luminosity distance of BBHs detected by each network at varying redshifts.
Adding Voyagers to the single ET network gives a small improvement in the distance estimation for nearby events. 
A negligible improvement is seen for intermediate-redshift events, and no improvement is demonstrated for faraway BBHs ($z > 6$).
While the \TGLE network performs similarly to the two-Voyager network for $z < 3$, a second ET offers more tangible improvement and allows more event distances to be constrained to within several hundred megaparsecs.
In most case, one deals with uncertainties of the order of \si$10$~Gpc.
Having a second 3G detector helps more for more distant events, since at $z\gtrsim 6$ the Voyager detectors will not contribute for the majority of sources.
However, even with these improvements, typical uncertainties will be of the order of $\sim 50$~Gpc.

We conclude this section by reminding that a measurement of BBHs position and distance will probably not be important in the context of an electromagnetic (EM) follow-up programs, as BBHs are not expected to produce any significant amount of EM radiation.
However, these measurements are important to verify if BBH sources are isotropically distributed and to study the large-scale structure of the Universe~\cite{2012PhRvD..86h3512J,2016PhRvD..94b4013N}, for some tests of general relativity~\cite{2016CQGra..33p5004C}, and for statistical measurements of the cosmological parameters~\cite{SchutzNature,2012PhRvD..86d3011D}.

\subsection{Spin}\label{sec:spin}

Next, we focus on the measurement of the spin of the black holes. A precise estimation of BH spins throughout cosmic history could help studying their formation channels and how their relative branching ratio evolved as a function of redshift.

We find that spins are one of the parameters for which estimation is least improved by the addition of Voyager detectors.

\begin{figure*}[htb!]
\includegraphics[width=\textwidth]{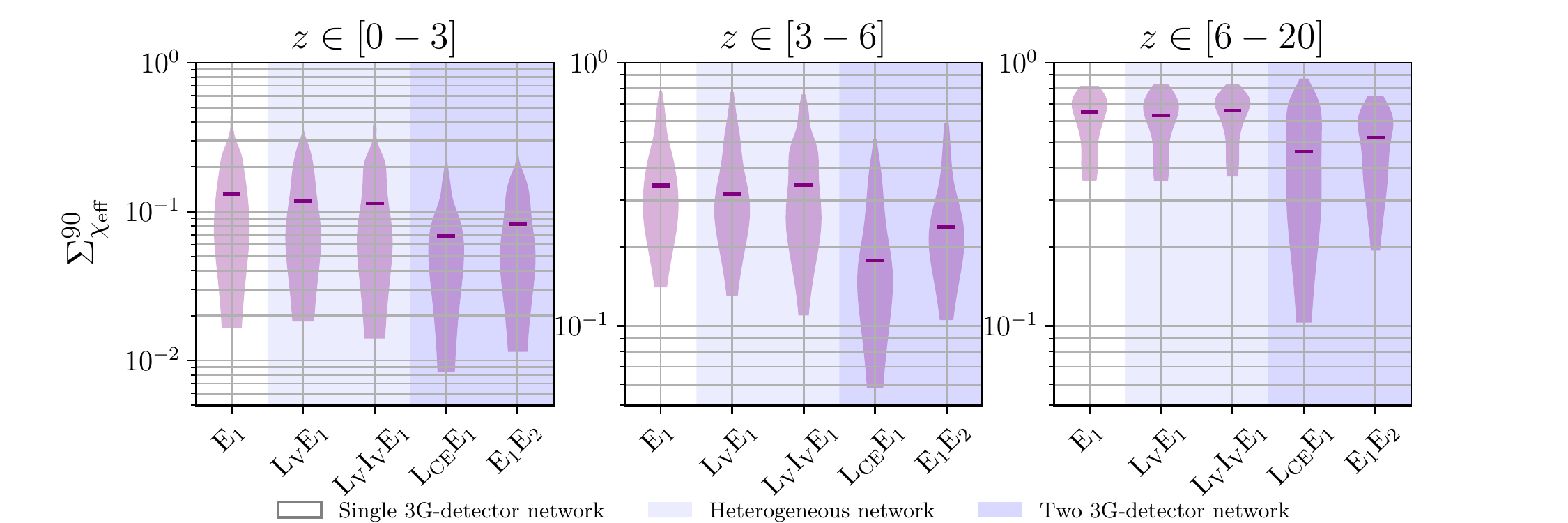}
\caption{Violin plots for the 90\% uncertainties for the effective spin parameter \chieff (y axis) obtained from each network (x axis). }
\label{fig:chieff}
\end{figure*}

Fig.~\ref{fig:chieff} shows the 90\% credible intervals calculated for the effective spin parameter \chieff~\cite{Damour:2001,Ajith:2009bn,Santamaria:2010}, which is the mass-weighted projection of the total spin along the orbital angular momentum vector. This combination can be typically be measured better than either of the component spins~\cite{Vitale:2016avz}. For advanced detectors, this will often be the \emph{only} spin parameter that can be measured~\cite{GW150914-DETECTION,GW151226-DETECTION,2017ApJ...851L..35A,2017PhRvL.119n1101A}.

On its own, a single ET can determine spins of average nearby BBHs to within \si$0.15$ 90\% confidence intervals, improved by about an order of magnitude for louder events.
In comparison, O1 saw \chieff constrained within intervals of 0.24, 0.5 and 0.30 for \Event{}, \Second{} and \Xmas{} respectively~\cite{2016PhRvX...6d1015A}.
In O2, GW170104, GW170608 (Hanford-Livingston) and GW170814 (Hanford-Livingston-Virgo) had 90\% confidence intervals of 0.51, 0.32 and 0.24 respectively~\cite{2017PhRvL.118v1101A,2017ApJ...851L..35A,2017PhRvL.119n1101A}.
We see that a single ET can therefore perform as well as, or better than, Advanced LIGO (+Virgo) in O1 and O2.
For BBHs at $3 < z < 6$, the \TGE network cannot estimate spins to better than \si$0.3$ on average.
For most faraway events, almost no information can be inferred about spins: for the networks with  a single 3G detector the median uncertainty is \si$0.65$ (the 90\% interval of the \emph{prior} on \chieff is $0.81$ using a prior uniform in the individual spins magnitude and directions, which is what has been done so far in all of LIGO-Virgo's papers~\cite{GW150914-DETECTION,GW151226-DETECTION,2017ApJ...851L..35A,2017PhRvL.119n1101A,2017PhRvL.118v1101A,2017PhRvL.119y1103V}).

The \TGLE network does show improvement in spin estimation by a small factor for nearby events.
A slight improvement is likewise evident in the \TGEE network. Even at redshifts of many, two-3G networks can occasionally yield \chieff constraints similar to those measured thus far with Advanced LIGO (+Virgo).

Networks with 3G detectors can often put significant constraints on the \emph{precessing} spin parameter \chip~\cite{Hannam:2013oca}, as well as on the \emph{component} spin magnitude, in particular that of the primary (i.e. most massive) body. Neither of these quantities has been measured by advanced detectors so far~\cite{GW150914-DETECTION,GW151226-DETECTION,2017ApJ...851L..35A,2017PhRvL.119n1101A}, and even as LIGO and Virgo progress toward design sensitivity, they will be only occasionally measurable~\cite{Vitale:2016avz}.

\begin{figure}[htb!]
\includegraphics[width=0.49\textwidth]{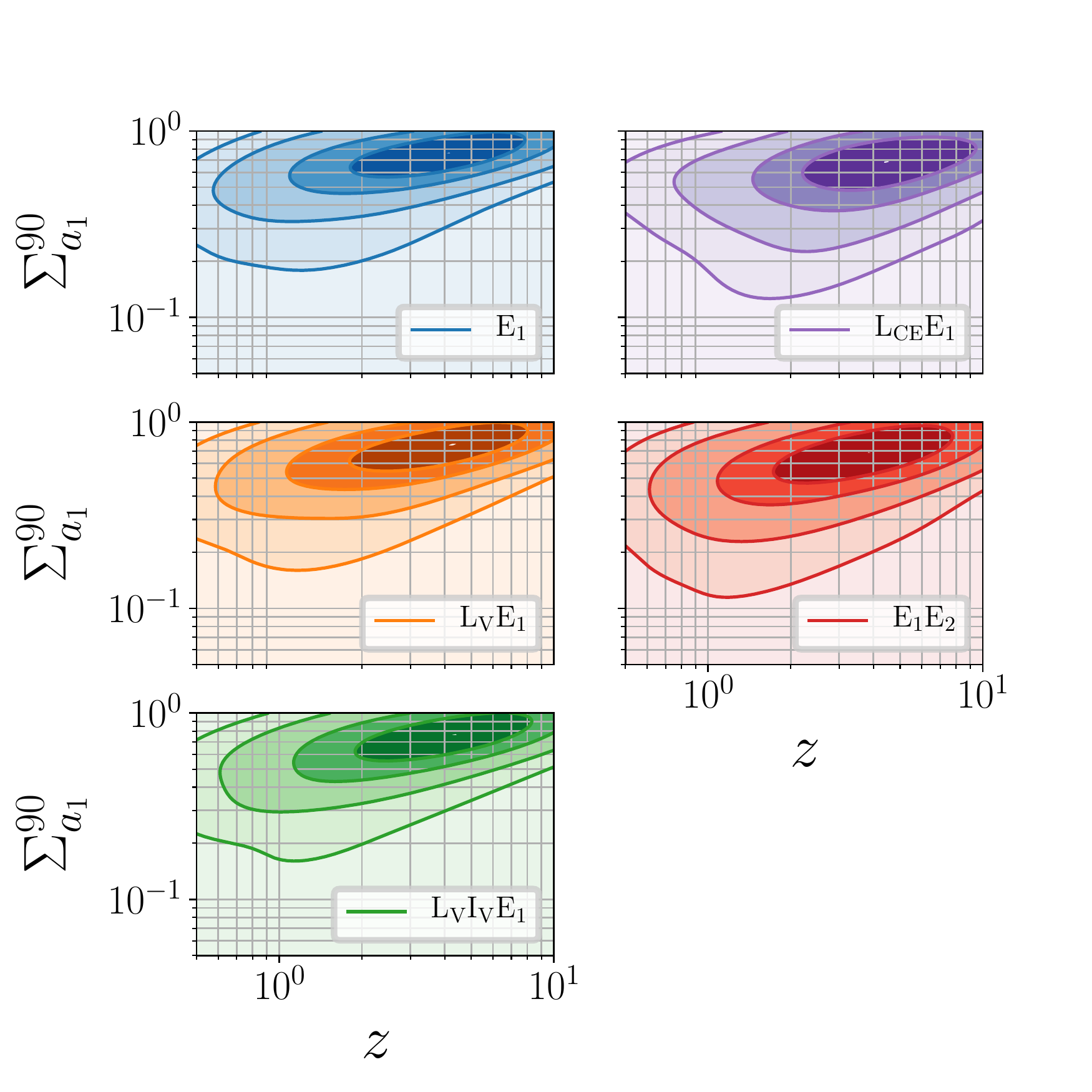}
\caption{Kernel-density estimate of the event distribution over primary spin $a_1$ as a function of redshift. The distribution was constructed using Gaussian kernels. Contours contain 90\%, 75\%, 50\% and 25\% of events.}
\label{fig:a1kernel}
\end{figure}

In Fig.~\ref{fig:a1kernel} we show a kernel-density estimate giving the distribution of events across redshift and uncertainty in spin magnitude for the primary object.
Gaussian kernels were used to approximate this distribution in logarithm-space of the parameters.
We recognize a slight advantage of the \TGEE and \TGLE networks in estimating spin for high SNR, very nearby events, with the possibility of measuring spins with uncertainties reaching toward \si~0.1.
At redshifts on the order of several, none of the networks are able to recover spins with significant constraints.
As expected, the Voyager detectors do not significantly change the distribution one would obtain with a \TGE detector.

\begin{figure*}[htb!]
\includegraphics[width=\textwidth]{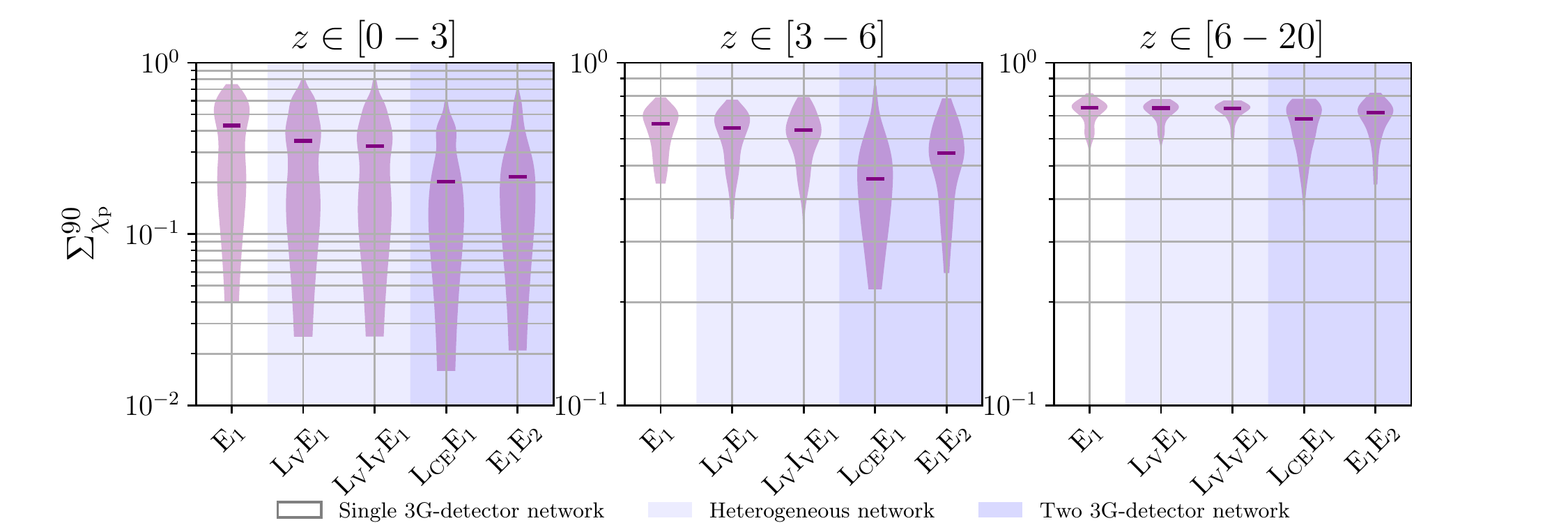}
\caption{Violin plots for the 90\% uncertainties for the precessing spin parameter \chip (y axis) obtained from each network (x axis).}
\label{fig:chip}
\end{figure*}

Finally, in Fig.~\ref{fig:chip} we show the uncertainty for the precessing spin parameter \chip. For $z<3$, networks with only one 3G detectors can estimate \chip for 50\% of the sources with uncertainties below $0.4$ (\TGE) or $0.3$ (\LE, \LEI). For two-3G networks, that number is 0.2. For reference, 90\% of the width of the \chip prior is 0.7. As mentioned above, for basically all of the events detected so far the posterior for \chip was extremely close to its prior~\cite{2017PhRvL.119y1103V,GW150914-DETECTION,GW151226-DETECTION,2017ApJ...851L..35A,2017PhRvL.119n1101A}.
As the redshift of the sources increases, only networks with two 3G detectors can still yield measurements of \chip with uncertainties below 0.5.

\subsection{Mass}

In this section we focus on the mass parameters. 

\begin{figure*}[htb!]
\includegraphics[width=\textwidth]{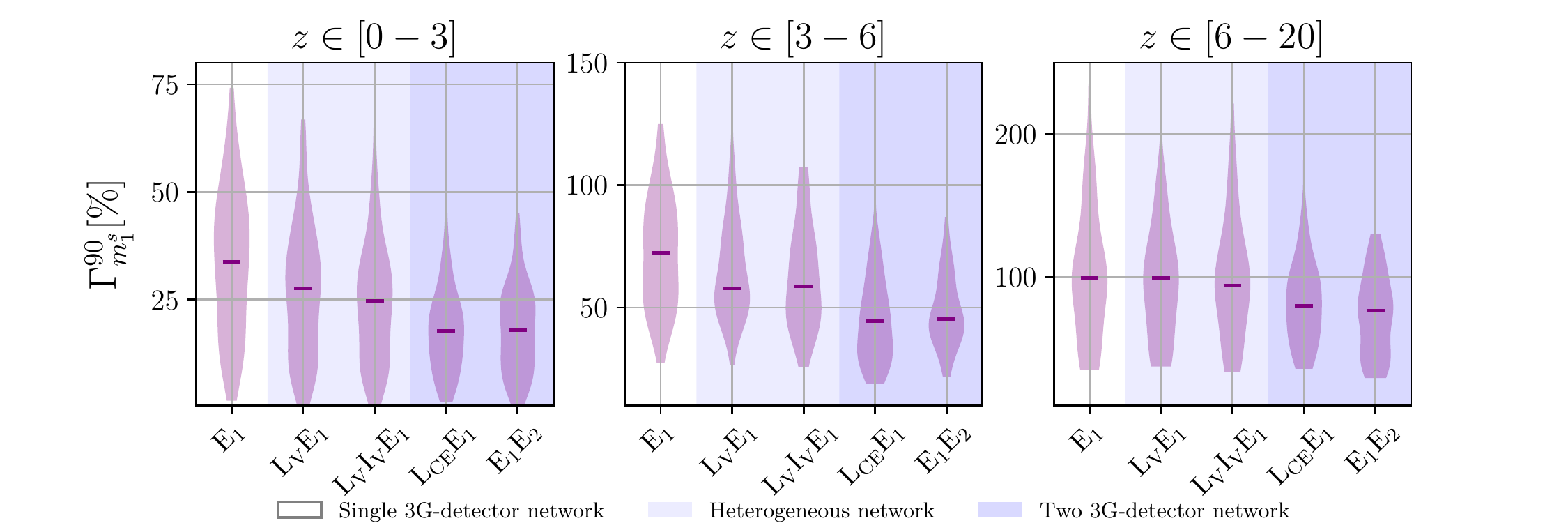}
\caption{Violin plots for the 90\% relative uncertainties for the source frame $m_1$ (y axis).}
\label{fig:m1_linear}
\end{figure*}

Fig.~\ref{fig:m1_linear} shows violin plots for the 90\% confidence interval for the source-frame mass of the primary object $m_1^s$.
For $0< z < 6$, the addition of Voyager(s) can significantly decrease the uncertainty achievable with a single  \TGE. This is not due to the extra SNR they provide (which is typically small, Fig.~\ref{fig:redshifts} top panel), but rather because they can help measuring the polarization of the signals, and hence the luminosity distance, Fig.~\ref{fig:distance}.
Since a measurement of the source's distance is required to infer the source-frame mass~\cite{Vitale3G}, the improvement in distance measurement corresponds to an improvement in mass measurement.
This translates to an improvement of the median estimation for the component masses by the \LEI network by a factor of \si1.5 over a single \TGE.

The addition of a second 3G detector further improves the mass estimation, yielding uncertainties below 50\% all the way to redshifts of many. 

\begin{figure*}[htb!]
\includegraphics[width=\textwidth]{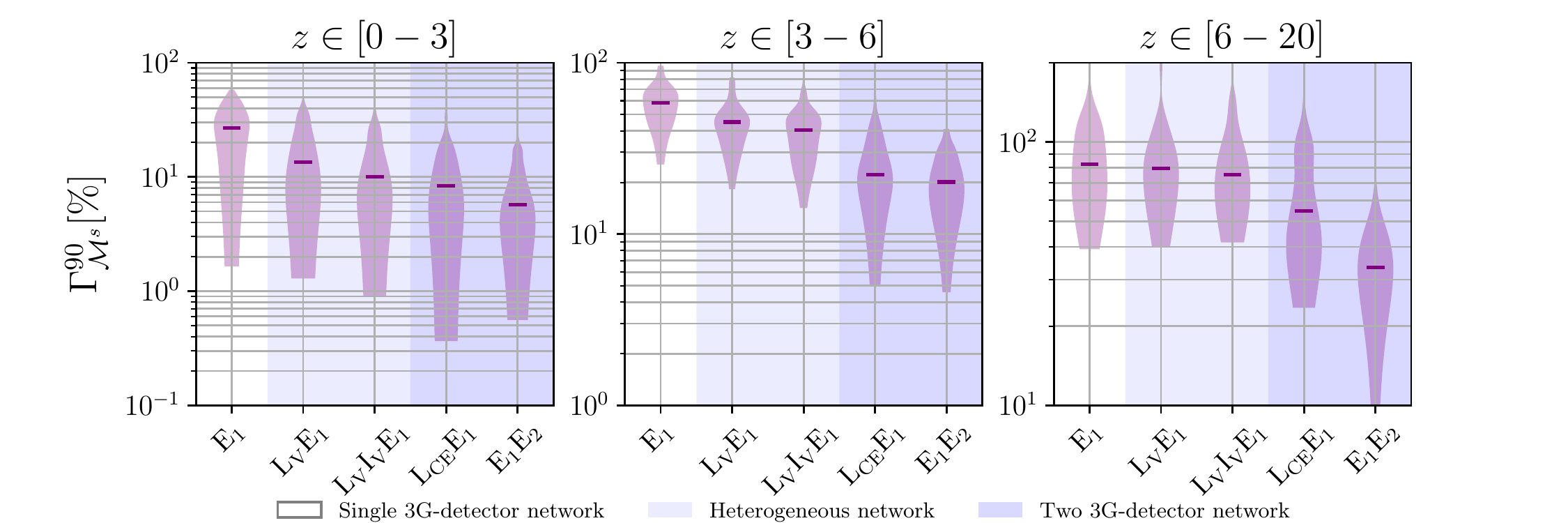}
\caption{Violin plots for the 90\% relative uncertainties for the source frame $\mathcal{M}^s$ (y axis).}
\label{fig:mc}
\end{figure*}

Similar trends are visible for the source-frame chirp mass,  $\mathcal{M}^s$, Fig.~\ref{fig:mc}. 
A single ET is only able to provide measurements more precise than 10\% for nearby BBHs. For comparison, the source-frame chirp mass of GW150914 was estimated with an uncertainty of $\sim 14\%$ (relative to the median)~\cite{GW150914-PARAMESTIM}.
For nearby events, each Voyager added results in a clear improvement. On the other end, at high redshift, the Voyager(s) are unable to provide additional polarization information and therefore give no improvement to the source-frame chirp mass uncertainty.

In the full 3G networks, we see significant increase in precision at all redshifts.
Of particular note is the ability of the \TGEE network to estimate the chirp mass to within \si$10\%$ for some events, even up to very high redshift.

\subsection{Other parameters}\label{Sec.Other}

A measurement of the orbital inclination for compact binaries which also emit EM radiation  can be useful to study the poorly known angular structure of the EM emission.
This was already the case for GRB 170817A that accompanied GW170817~\cite{2017ApJ...848L..13A}, and which appeared to be significantly under-luminous when compared to other GRBs.

While BBH are not expected to emit EM radiation, there still is interest in a precise estimation of the orbital inclination.
For example, tests of general relativity based on the measurement of the ringdown of the newly formed black holes improve with a more precise measurement of the inclination.
In these tests, one checks whether the ratios of the amplitudes and the decay times of ringdown quasinormal modes match what is prescribed by general relativity~\cite{PhysRevD.76.104044}.
Critically, theses ratios also depend on the inclination angle, which implies that a poorly measured inclination adds uncertainty to the test. Conversely, if the inclination angle is precisely measured, the uncertainties on the amplitude ratios are reduced. 

\begin{figure*}[htb!]
\includegraphics[width=\textwidth]{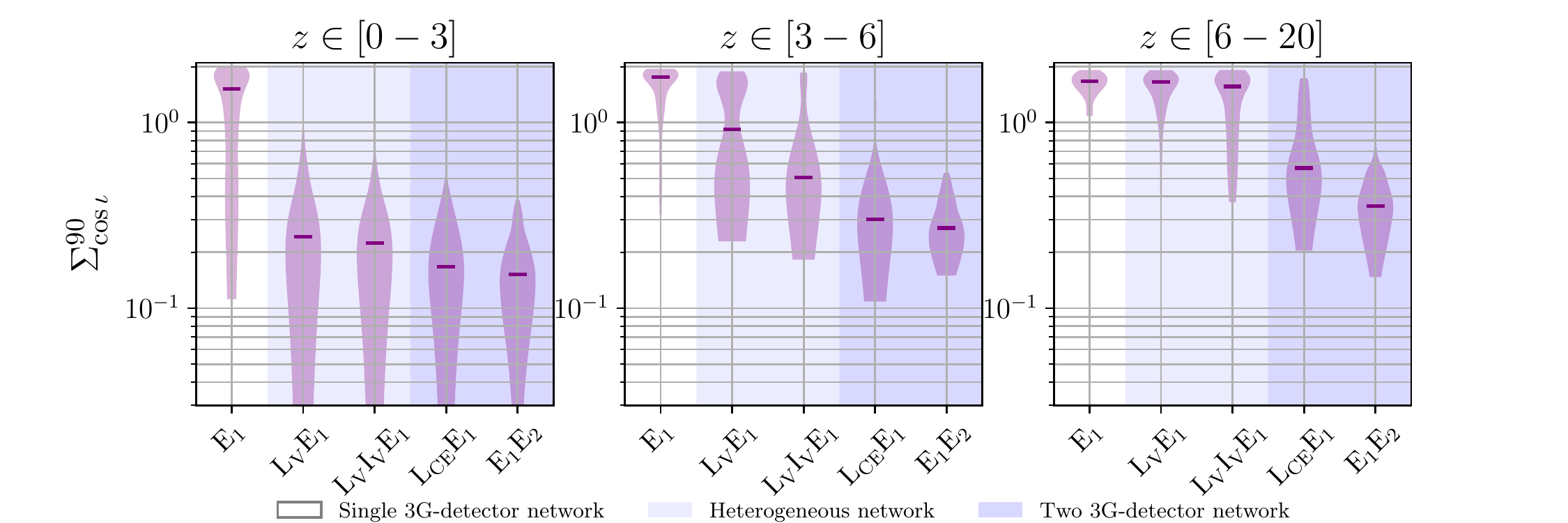}
\caption{Violin plots for the 90\% uncertainties for the cosine of the inclination angle $\cos \iota$ (y axis).}
\label{fig:cosiota}
\end{figure*}

Fig.~\ref{fig:cosiota} shows the 90\% confidence intervals for the cosine of the source inclination angle. 
We see that a single ET detector gives basically no information about the inclination of the source at all distances (aside from particularly loud events).
A single Voyager reduces the uncertainty by one order of magnitude. For the average event, a \LE network gives an uncertainty of \si$0.25$.

For sources at $3 < z < 6$, we still get some visible improvement with a heterogeneous network, whereas an additional 3G detector would give a significant boost in precision.
At high redshift, the Voyager detectors are no longer useful in narrowing down the source inclination, while a second CE or ET would yield improvements of the order of a few.

Similarly, the inclusion of more detectors provides more information about the polarization angle of the gravitational-wave signal $\psi$~\cite{1987thyg.book.....H}. Measurement of $\psi$ could facilitate some tests of general relativity, notably the detection of gravitational-wave memory~\cite{2016PhRvL.117f1102L}.

While we won't quote 90\% confidence intervals for the polarization angle, since it is very often multimodal, we note that Voyager-class detectors can significantly improve its measurability compared with a single ET.

\section{Golden events}\label{sec:nearby}

In the previous sections, we have shown how having previous-generation detectors can improve the measurement of the extrinsic parameters for most detectable systems. 
However, since the distances of GW sources are expected to be uniform in comoving volume, they will typically correspond to redshifts of the order of unity. 
In fact, none of the \si 210 events we randomly generated as described in the previous section had redshift smaller than $\sim 0.3$.

In this section we want to explicitly check how well heterogeneous and 3G networks will measure the parameters of local sources, such as those detected by the LIGO and Virgo detectors in their first two science runs.
While the bulk of the events detected by heterogeneous or 3G networks will live at redshift of the order of unity, we know that events at redshifts of \si 0.1 will be detectable often enough, and with signal-to-noise ratios large enough to dramatically contribute to the science output of future ground based detectors.

To facilitate comparison with existing results, we have created software replicas of GW150914 and GW151226 and added them to simulated interferometric noise, using all of the networks used above, as well as a single CE site.

The parameters of these simulated sources were selected randomly from the published 90\% credible intervals~\cite{O1-BBH}. In Tab.~\ref{Tab.NearbyInj} we report the true values of the main parameters as well as the SNR in each network.

\begin{table*}[htb]
\centering
\caption{Injected parameters for the software replica of GW150914 and GW151226. The chirp mass is the detector-frame. The last 6 columns report the optimal signal-to-noise ratio in each of the the networks we considered, as well as the SNR with which GW150914 and GW151226 were actually detected by Advanced LIGO~\cite{O1-BBH}}
\label{Tab.NearbyInj}
\begin{tabular}{|c|c|c|c|c|c|c|c|c|c|c|c|c|c|}
\hline
 & $\mathcal{M}_\mathrm{det}\;[M_\odot]$ & q     & $\chi_\mathrm{eff}$ & $\chi_\mathrm{p}$ & $D_L$ [Mpc] & $\cos\theta_\mathrm{JN} $& $\rho_{\TGE}$ & $\rho_{\LE}$ & $\rho_{\LEI}$ & $\rho_{\TGL}$ & $\rho_{\TGLE}$& $\rho_{\TGEE}$& $\rho_{\mathrm{Adv.Det}}$~\cite{O1-BBH} \\ \hline
GW150914 & 29.95                                 & 0.964 & -0.124              & 0.559             & 497.01   &  0.971 & 687      & 712         & 746          & 1965                   & 2081      &  1210 & 23.7           \\ \hline
GW151226 & 9.672                                 & 0.833 & 0.135               & 0.574             & 573.81     & 0.898& 199      & 225         & 244          & 1087                   & 1106      &  408  &  13.0       \\ \hline
\end{tabular}
\end{table*}

A few points are worth stressing. 
First, for both events, the SNR contributed by one or two Voyager is negligible when compared to the SNR in ET. However, the SNR in each of the Voyager will be well above detection threshold, of the order of several tens. 
To a lesser extent, the same can be said of the SNR contributed by ET in a \TGLE network, where the large majority of the network SNR comes from the CE.
The fact that the SNR in the least sensitive detector is well above threshold implies that those detectors can still significantly contribute to the measurement of the source's extrinsic parameters: sky position and distance.
On the other hand, the measurement of intrinsic parameters such as the (detector frame) chirp mass and the spins will be dominated by the best detector in the network, and only marginally improve as less sensitive instruments are added.

This is confirmed by the results presented in Tab.~\ref{Tab.NearbyGW150914} for GW150914 and Tab.~\ref{Tab.NearbyGW151226} for GW151226.
In each of these tables, we report the size of the 90\% credible interval for a few key parameters, for all the networks we considered, as well as the uncertainties reported by Ref.~\cite{O1-BBH} for the actual detections.

\begin{figure*}[htb]
\begin{tabular}[c]{cc}
\centering
\begin{subfigure}{0.4\textwidth}
\includegraphics[width=\textwidth]{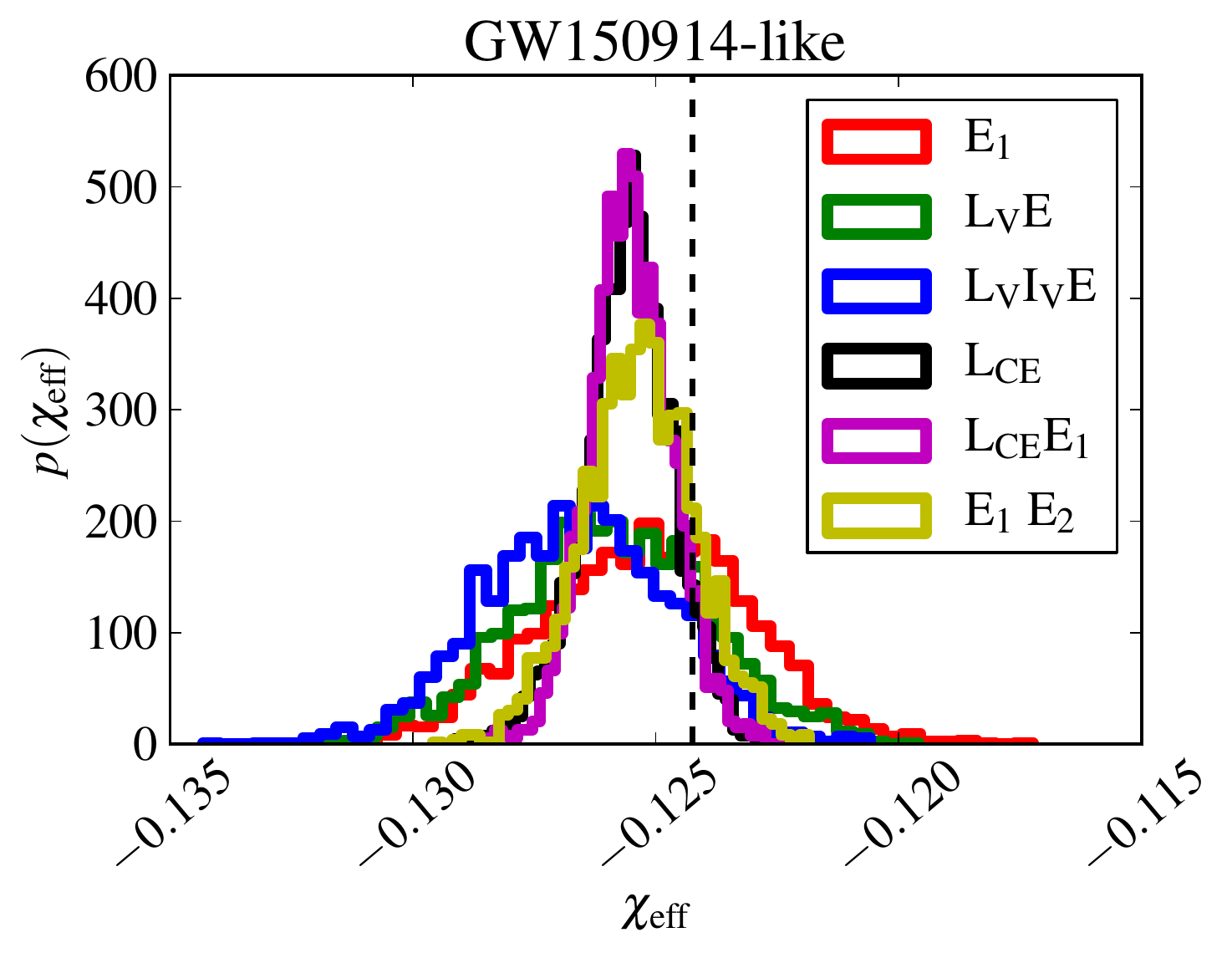}
\end{subfigure} & 
\begin{subfigure}{0.4\textwidth}
\includegraphics[width=\textwidth]{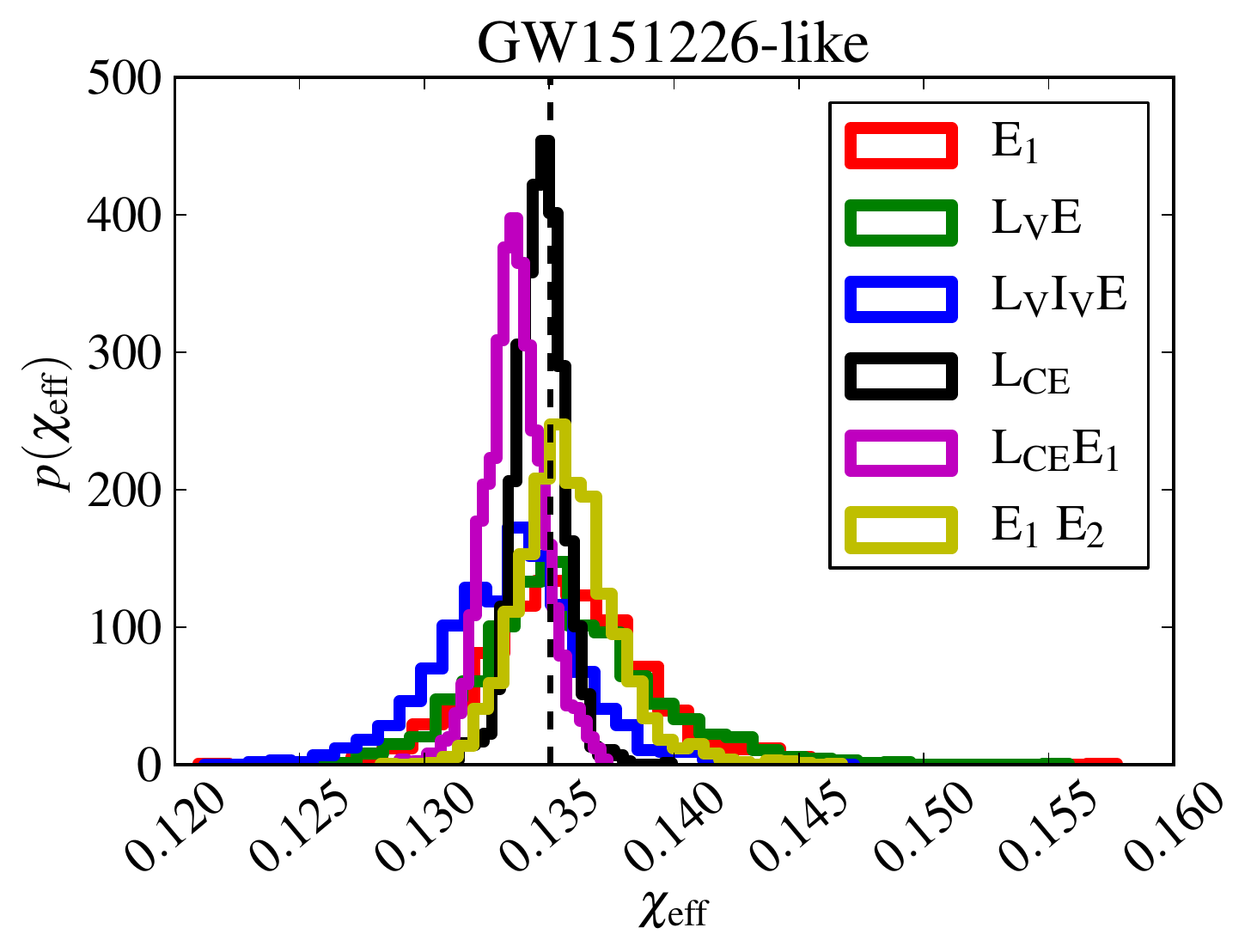}
\end{subfigure} \\
\begin{subfigure}{0.4\textwidth}
\includegraphics[width=\textwidth]{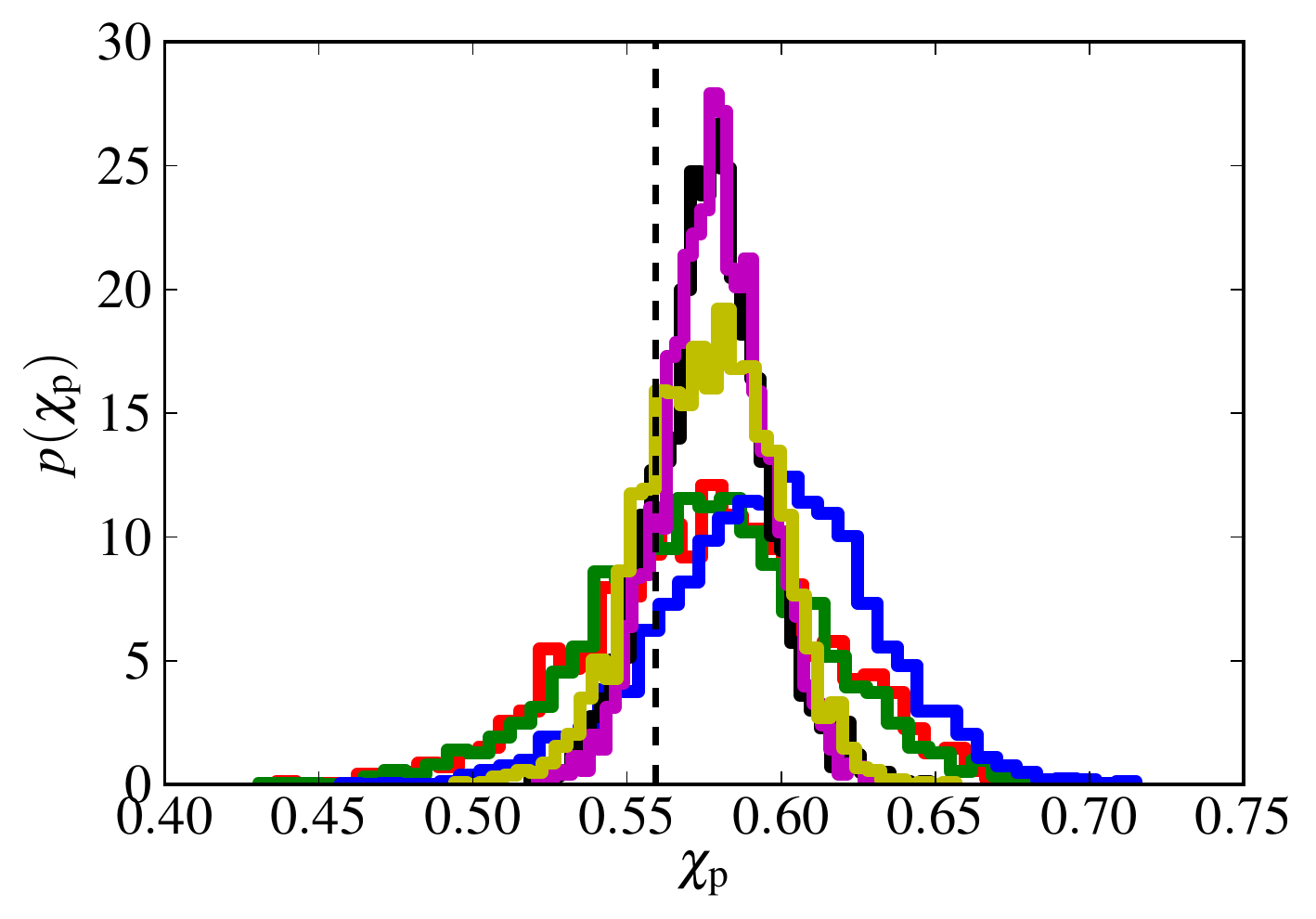}
\end{subfigure} & 
\begin{subfigure}{0.4\textwidth}
\includegraphics[width=\textwidth]{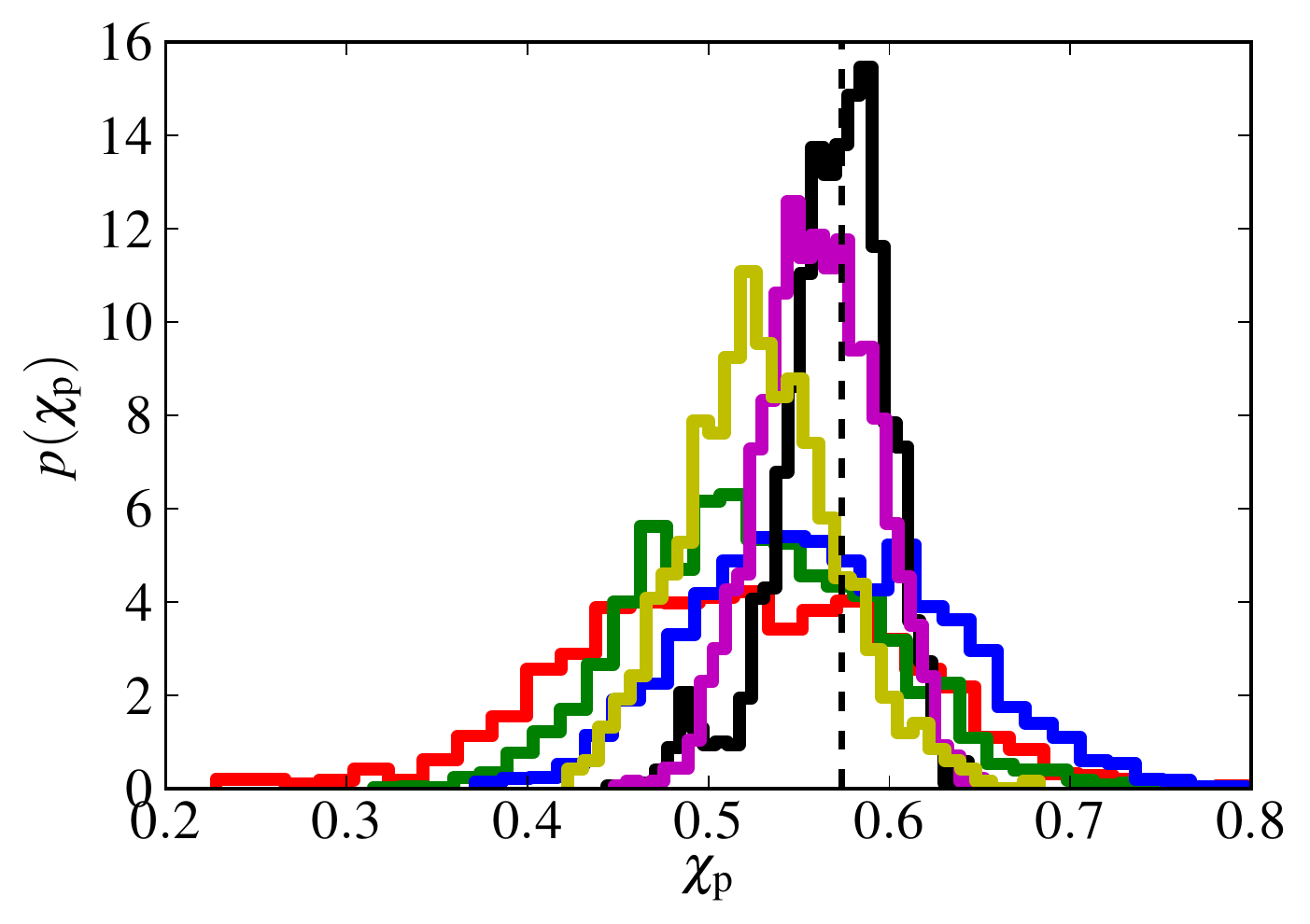}
\end{subfigure} \\
\begin{subfigure}{0.4\textwidth}
\includegraphics[width=\textwidth]{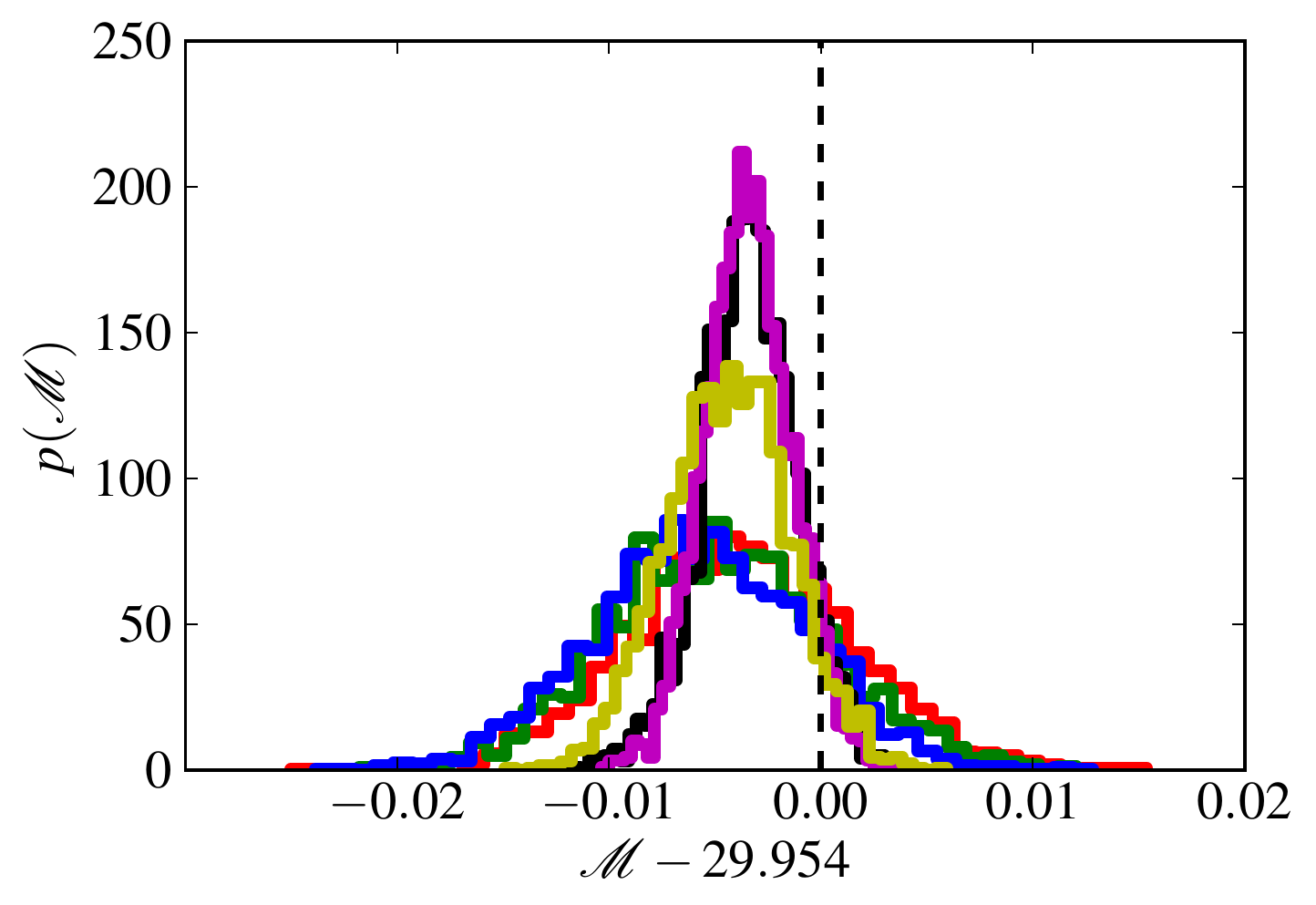}
\end{subfigure} & 
\begin{subfigure}{0.4\textwidth}
\includegraphics[width=\textwidth]{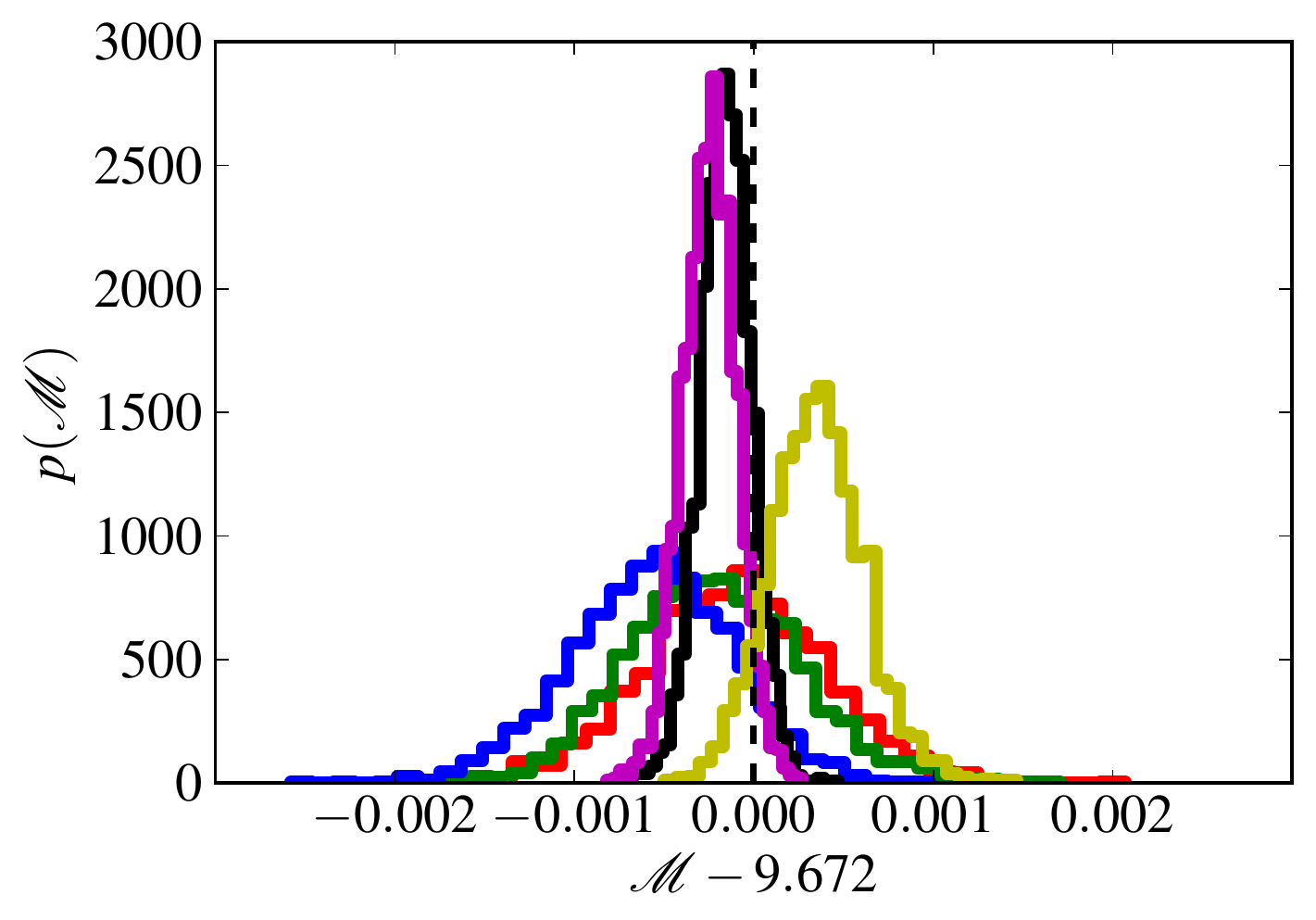}
\end{subfigure} \\
\begin{subfigure}{0.4\textwidth}
\includegraphics[width=\textwidth]{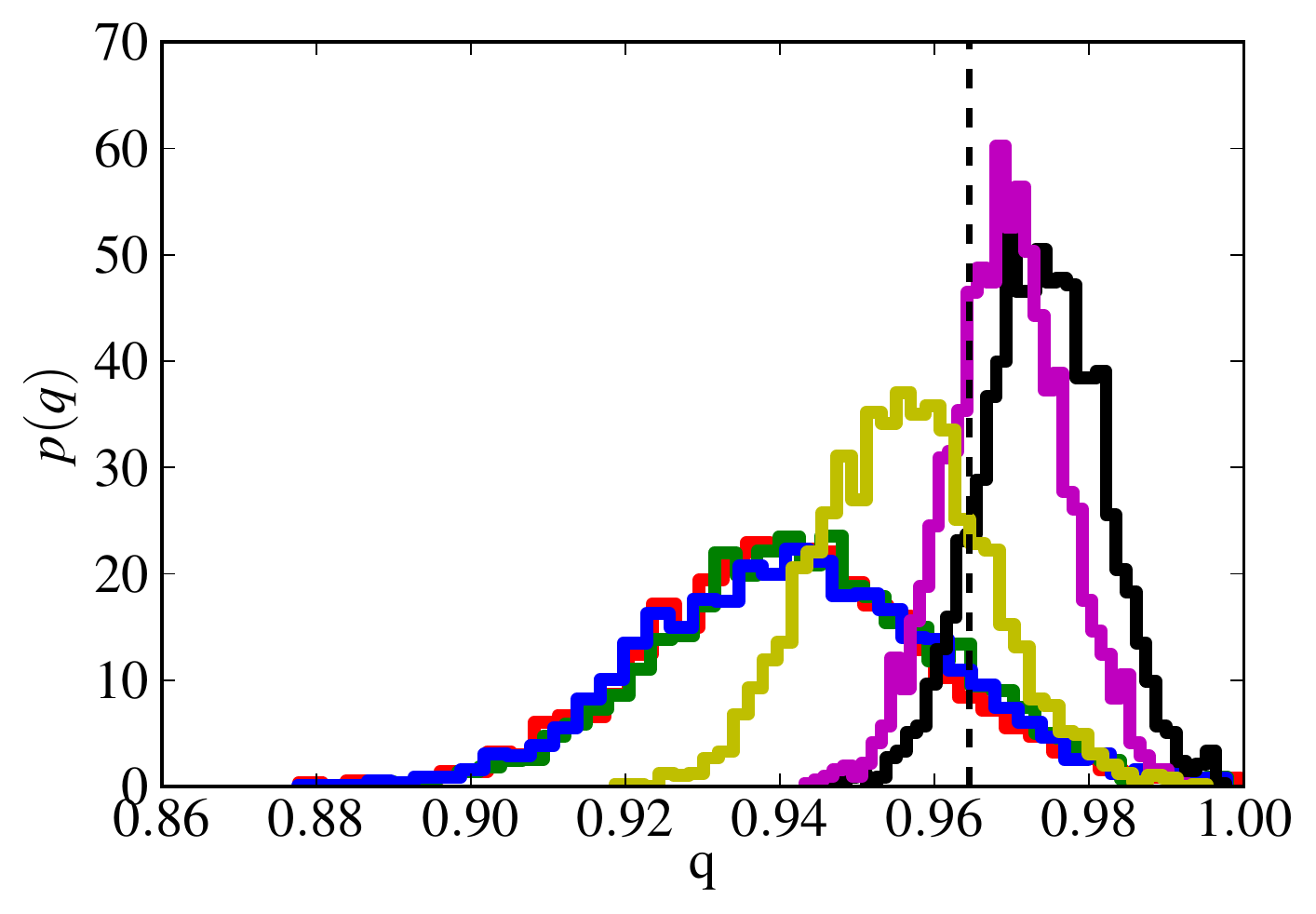}
\end{subfigure} & 
\begin{subfigure}{0.4\textwidth}
\includegraphics[width=\textwidth]{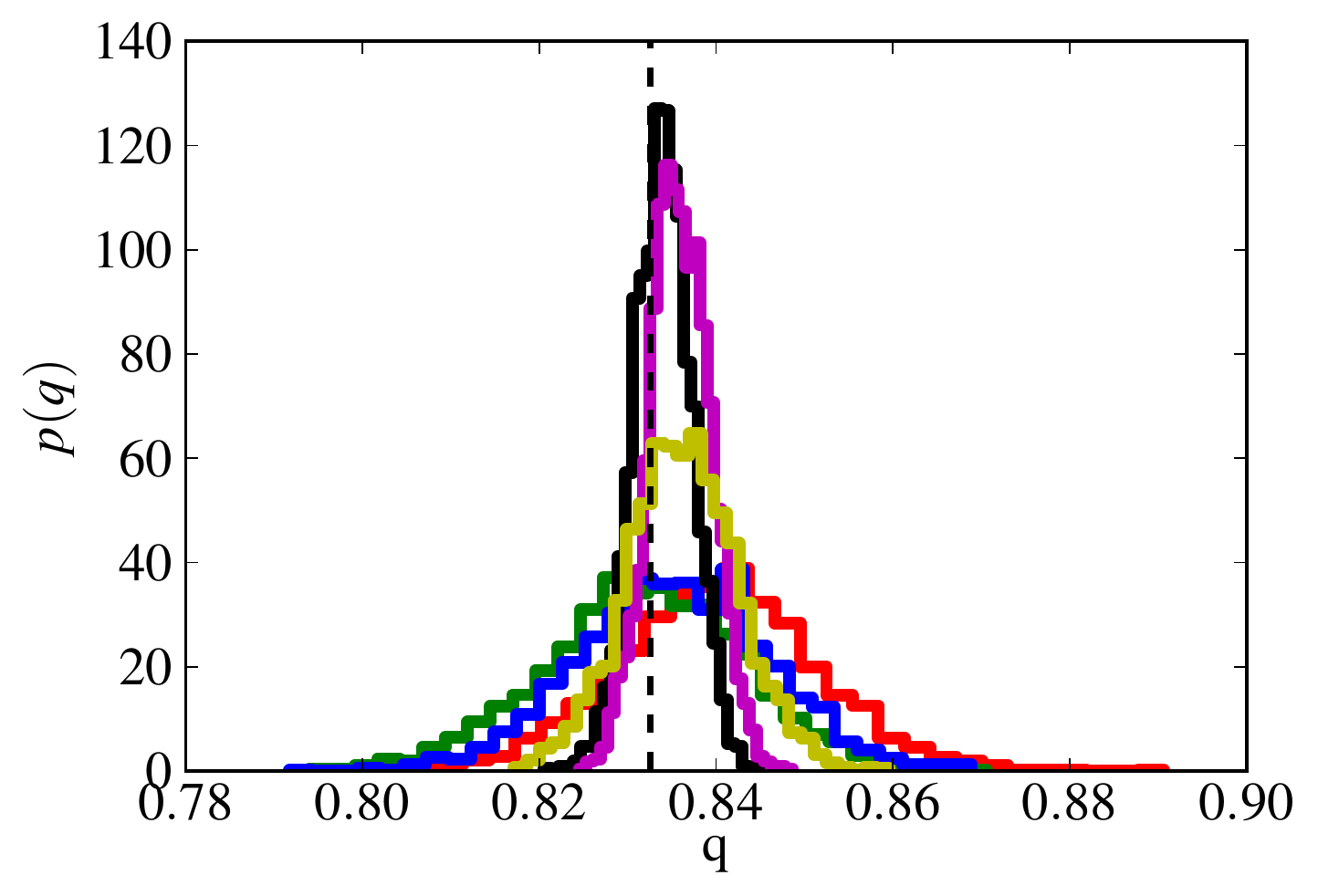}
\end{subfigure} \\
\end{tabular}\caption{Posterior distributions for the intrinsic parameters of  a GW150914-like signal (left column) and a GW151226-like signal (right column) as detected by 3G and heterogeneous networks. A dashed line marks the true value of the parameters.}\label{Fig.ReplicaInt}
\end{figure*}

\begin{figure*}[htb]
\begin{tabular}[c]{cc}
\centering
\begin{subfigure}{0.4\textwidth}
\includegraphics[width=\textwidth]{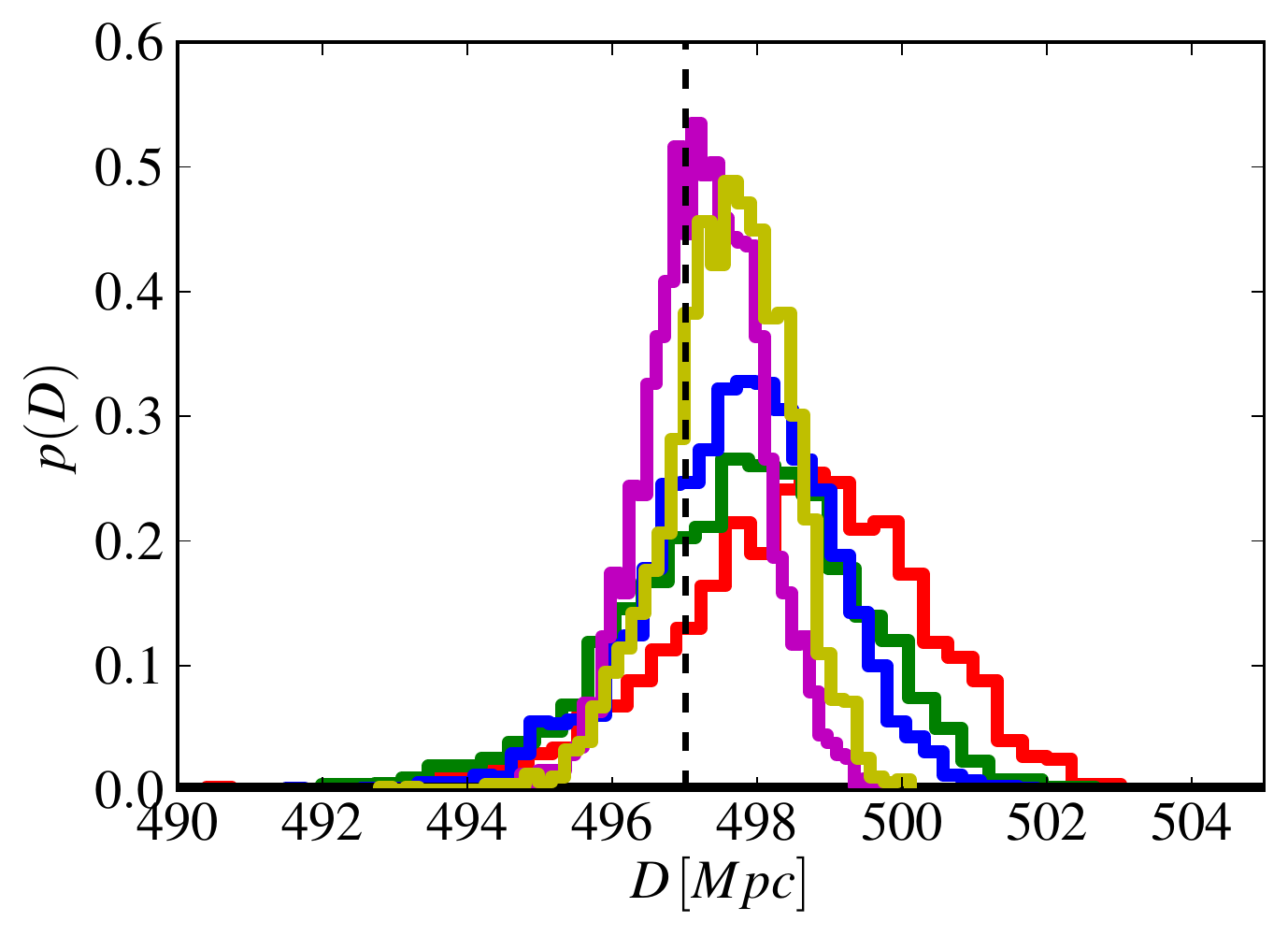}
\end{subfigure} & 
\begin{subfigure}{0.4\textwidth}
\includegraphics[width=\textwidth]{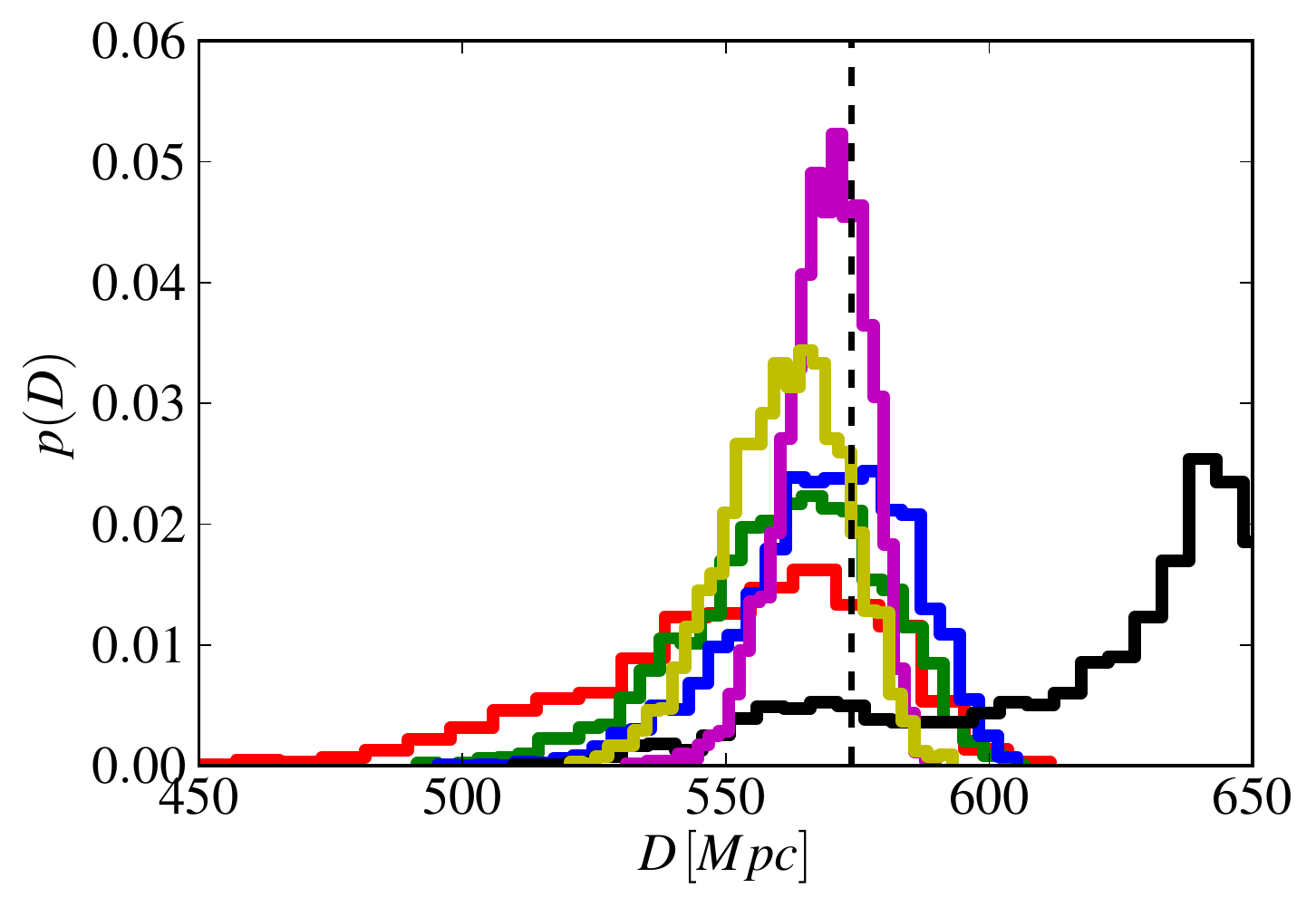}
\end{subfigure} \\
\begin{subfigure}{0.4\textwidth}
\includegraphics[width=\textwidth]{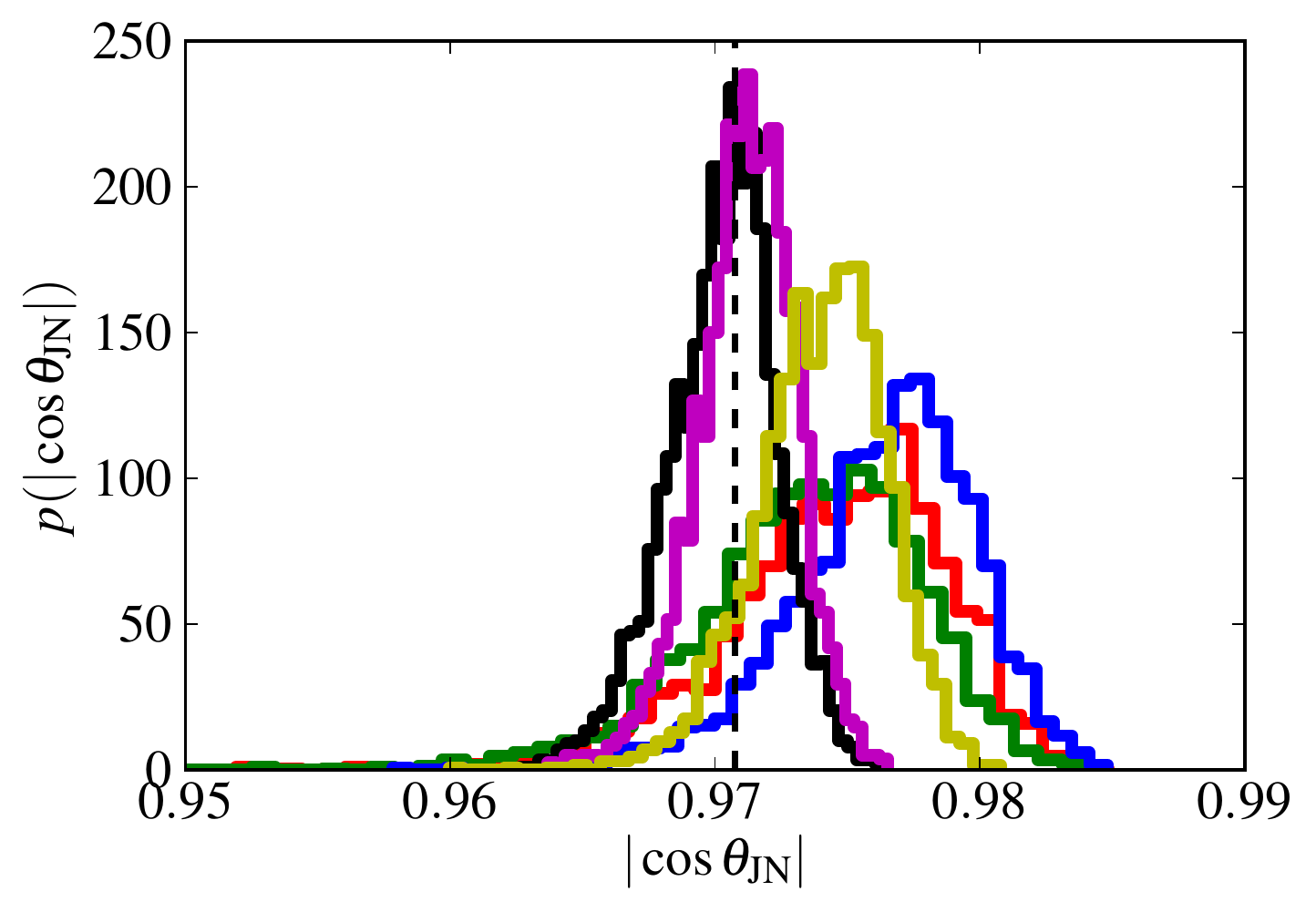}
\end{subfigure} & 
\begin{subfigure}{0.4\textwidth}
\includegraphics[width=\textwidth]{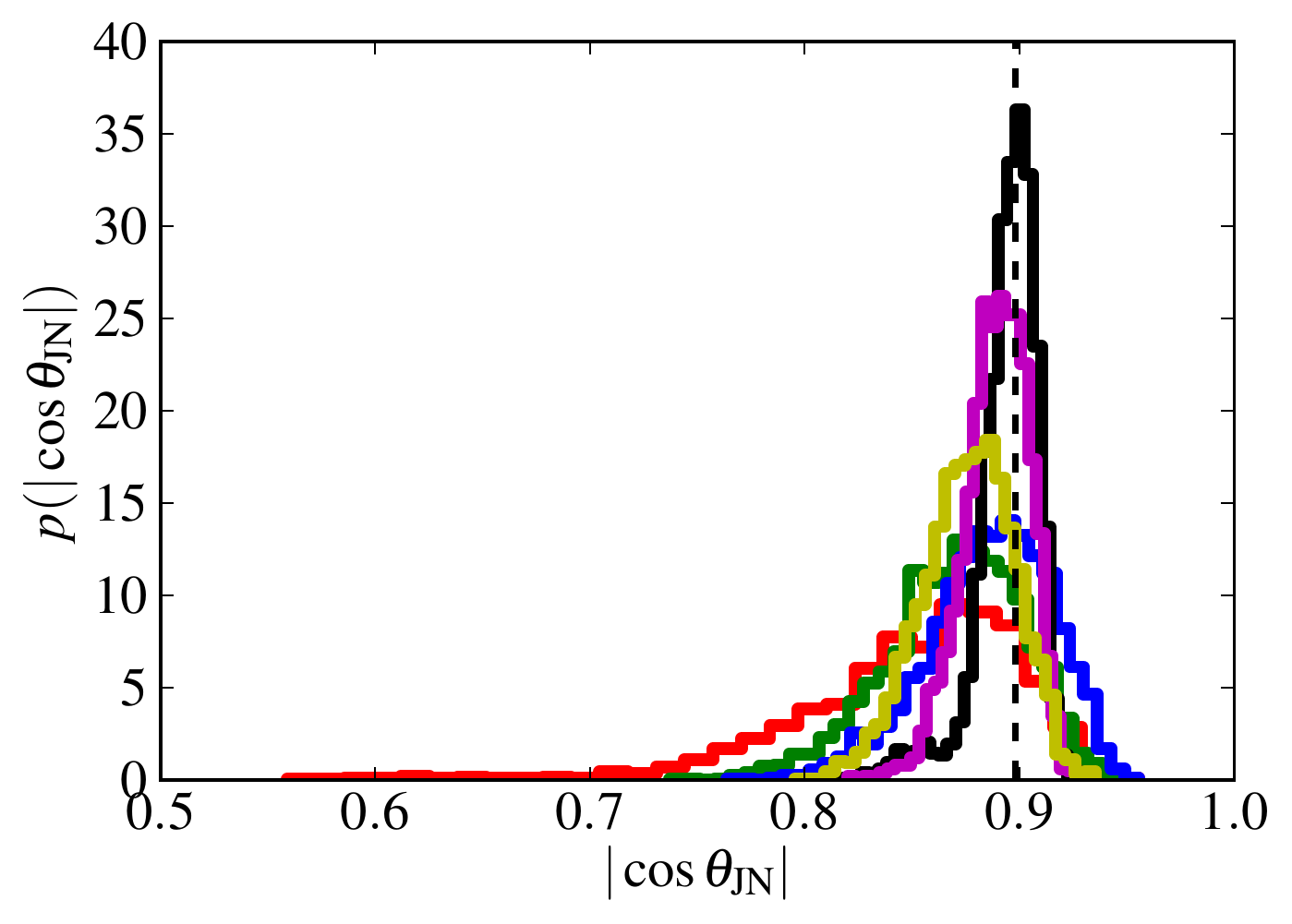}
\end{subfigure} \\
\end{tabular}\caption{Posterior distributions for the extrinsic parameters of a GW150914-like signal (left column) and a GW151226-like signal (right column) as detected by 3G and heterogeneous networks. A dashed line marks the true value of the parameters.}\label{Fig.ReplicaExt}
\end{figure*}

We will focus on GW150914, Tab.~\ref{Tab.NearbyGW150914} and left panels of Fig.~\ref{Fig.ReplicaInt} (mass and spins) and Fig.~\ref{Fig.ReplicaExt} (distance, sky position and inclination), as similar trends are visible for the GW151226 replica.
We find that the uncertainty in the measurement of the (detector frame) chirp mass is of the order of $10^{-2} M_\odot$ for \TGE, \LE, \LEI.  The uncertainty goes down by a factor of \si 2 if a CE is added to the network, or a factor of \si~1.5 with a two-ET network. For comparison, Advanced LIGO measured the detector frame chirp mass of GW150914 with an uncertainty of $3 M_\odot$.
Similar trends are present for the mass ratio, $q$, which can measured with uncertainties $\sim 6\times 10^{-2}$ by \TGE or the heterogeneous networks. As expected, the results from the heterogeneous networks do not significantly improve w.r.t.\ what one could measure with ET alone. When \TGEE is used, the SNR increases, yielding an uncertainty of \si$3.6\times10^{-2}$. Finally, the higher sensitivity of CE allows for uncertainties as small as $\sim 2.5\times 10^{-2}$. The numbers should be compared to the 90\% credible interval of $0.34$ for Advanced LIGO~\cite{O1-BBH}.
Fig.~\ref{Fig.ReplicaInt} shows how the posteriors of \TGE, \TGEE and \TGL can exclude equal mass ($q=1$) at high confidence level, even for a system where the true mass ratio is extremely close to unity.

Similar trends are visible for the spin parameters. Here too, adding one or two Voyager to a pre-existing ET does not significantly change the results; whereas a factor of $\sim 2$ is gained by adding another 3G detector. 
It is worth stressing that even a single 3G site could yield an actual measurement of the precessing spin parameter \chip, whereas for all events detected so far by advanced detectors the posterior for \chip was not dissimilar from the prior~\cite{GW150914-PARAMESTIM,GW151226-DETECTION,2017PhRvL.119y1103V,2017ApJ...851L..35A,2017PhRvL.119n1101A,2017PhRvL.118v1101A}.
In Fig.~\ref{Fig.ReplicaInt}, left column, we show the posterior distribution for the intrinsic parameters of the GW15014-like event, as measured by all networks (we do not show the uncertainties reported in Ref.~\cite{O1-BBH} as those would require much larger ranges for the x axis).
For extrinsic parameters, having Voyager detectors alongside a single ET can make a dramatic difference. 

In the bottom panels of Fig.~\ref{Fig.ReplicaExt} we show the uncertainties of the absolute value of the inclination angle, $\theta_\mathrm{JN}$. We calculate the absolute value because the posteriors are bimodal for the single-detector networks \TGL and \TGE, which would make the plot hard to read. The uncertainties in Tab.~\ref{Tab.NearbyGW150914} instead do take the sign of the posterior into account. As mentioned in Sec.~\ref{Sec.Other}, precise inclination measurement could help for some tests of general relativity.

Still in Tab.~\ref{Tab.NearbyGW150914} we see that the sky localization of a GW150914-like BBH would be extremely poor with a single L-shaped 3G detector. Owing to its triangular geometry, a single ET would do better, localizing the source to within $0.16$ deg$^2$. 
If  a single Voyager is added to ET, the uncertainty goes down by nearly 1 order of magnitude, to $2.1\times 10^{-2}$ deg$^2$; and another factor of 2 can be gained adding a second Voyager.
Networks of 2-3G instruments can localize to areas of few~$\times10^{-3}$~\degg, with \TGEE better than \TGLE due to its geometry.

\begin{table*}[htb]
\centering
\caption{90\% credible intervals for various parameters for a software replica of GW150914, as detected by the future networks considered in this study. For reference, we give the 90\% credible interval published by the LIGO and Virgo collaborations in Ref.~\cite{O1-BBH}.}\label{Tab.NearbyGW150914}
\begin{tabular}{|c|c|c|c|c|c|c|c|}
\hline
&E$_1$&L$_{\mathrm{V}}$E&L$_{\mathrm{V}}$I$_{\mathrm{V}}$E&L$_{\mathrm{CE}}$&L$_{\mathrm{CE}}$E$_1$&E$_1$ E$_2$&Adv. Det\\ \hline$\mathcal{M} \,[\mathrm{M}_\odot]$& 1.7e-02&1.7e-02&1.6e-02&7.6e-03&6.8e-03&9.5e-03&3.0e+00\\ \hline$q $& 6.1e-02&6.0e-02&6.1e-02&2.5e-02&2.5e-02&3.6e-02&3.4e-01\\ \hline$\chi_{\mathrm{eff}} $& 6.8e-03&6.5e-03&6.2e-03&3.1e-03&2.8e-03&3.8e-03&2.3e-01\\ \hline$\chi_{\mathrm{p}} $& 1.3e-01&1.2e-01&1.1e-01&5.8e-02&5.5e-02&6.9e-02&-\\ \hline$D \,[\mathrm{Mpc}]$& 5.6e+00&5.2e+00&4.2e+00&3.4e+02&2.5e+00&2.8e+00&3.1e+02\\ \hline$\cos{{\theta}_{\mathrm{JN}}}$& 2.0e+00&1.3e-02&1.1e-02&1.9e+00&6.0e-03&7.8e-03&-\\ \hline$\delta \Omega \,[\mathrm{Deg}^2]$& 1.6e-01&2.1e-02&1.0e-02&2.8e+03&5.9e-03&2.5e-03&2.3e+02\\
\hline
\end{tabular}
\end{table*}

For the luminosity distance (Fig.~\ref{Fig.ReplicaExt}, second panel from the bottom) we find similar trends, although the improvements from adding a Voyager are smaller. The measurement of the luminosity distance from a single CE (\TGL) also yields a very poor result due to the impossibility of measuring both polarizations of a GW with a single L-shaped detector.
Since the measurement of the \emph{source-frame} mass parameters requires a measurement of the luminosity distance~\cite{Vitale3G}, the smaller uncertainties in the measurement of the \emph{detector-frame} masses by \TGLE compared to \LEI might not directly translate to a better estimation of the astrophysically relevant source-frame parameters.

\begin{table*}[]
\centering
\caption{90\% credible intervals for various parameters for a software replica of GW151226, as detected by the future networks considered in this study. For reference, we give the 90\% credible interval published by the LIGO and Virgo collaborations in Ref.~\cite{O1-BBH}.}
\label{Tab.NearbyGW151226}
\begin{tabular}{|c|c|c|c|c|c|c|c|}
\hline
&E$_1$&L$_{\mathrm{V}}$E&L$_{\mathrm{V}}$I$_{\mathrm{V}}$E&L$_{\mathrm{CE}}$&L$_{\mathrm{CE}}$E$_1$&E$_1$ E$_2$&Adv. Det\\ \hline$\mathcal{M} \,[\mathrm{M}_\odot]$& 1.7e-03&1.6e-03&1.5e-03&4.6e-04&5.0e-04&8.7e-04&1.2e-01\\ \hline$q $& 3.7e-02&3.7e-02&3.5e-02&1.1e-02&1.1e-02&2.1e-02&6.4e-01\\ \hline$\chi_{\mathrm{eff}} $& 1.1e-02&1.2e-02&9.2e-03&3.0e-03&3.7e-03&6.2e-03&2.3e-01\\ \hline$\chi_{\mathrm{p}} $& 2.8e-01&2.1e-01&2.3e-01&9.4e-02&1.0e-01&1.4e-01&-\\ \hline$D \,[\mathrm{Mpc}]$& 9.3e+01&5.9e+01&5.4e+01&1.0e+02&2.6e+01&3.8e+01&3.5e+02\\ \hline$\cos{{\theta}_{\mathrm{JN}}}$& 1.8e+00&1.0e-01&9.5e-02&4.5e-02&4.9e-02&7.1e-02&-\\ \hline$\delta \Omega \,[\mathrm{Deg}^2]$& 1.9e+00&1.4e-01&5.3e-02&1.2e+03&5.8e-02&5.9e-03&8.5e+02\\
\hline
\end{tabular}
\end{table*}
We stress how the sky and distance uncertainties imply that the 3D localization error volume would contain on average one  galaxy for all networks except \TGL.
This can be easily seen using a reasonable estimation of galaxy density, $0.01~\mathrm{galaxy/Mpc}^3$~\cite{1976ApJ...203..297S,2016arXiv161201471C}, and calculating the 3D volume corresponding to the uncertainties given in Table.~\ref{Tab.NearbyGW150914}.
This implies that an adequate EM follow-up should be able to uniquely identify the host galaxy to the BBH merger even in absence of any EM signals. Then, the luminosity distance of the GW source and the redshift measured from the host galaxy can be used to infer the Hubble constant~\cite{SchutzNature}. If the redshift of the source due to expansion of the Universe were perfectly known, the relative uncertainty in the luminosity distance would translate in the same relative uncertainty in the Hubble constant~\cite{2017Natur.551...85A}. In practice, the redshift estimation is affected by photometric and/or spectroscopic uncertainty, and by the peculiar velocity of the host galaxy relative to the Hubble flow~\cite{2017Natur.551...85A}.

Using the BBH merger rates calculated by the LVC, one can estimate that $10^5$ BBH merge each year~\cite{2017PhRvL.118o1105R}. If one assumes those have redshift uniform in comoving volume, it follows that roughly $200$ BBH would have uncertainties similar to or better than the ones we presented in Table~\ref{Tab.NearbyGW150914} for each year of observation. 
For our GW150914-replica the uncertainty in the luminosity distance is on the order of \si0.5\% for the best networks.
As the uncertainty given $N$ detections scales roughly as $1/\sqrt{N}$~\cite{2017arXiv171206531C}, BBH sources can potentially yield a measurement of the Hubble constant with \si 0.03\% precision after one year of data. This number does not include the uncertainty on the redshift, and on the calibration of future ground-based detectors, both of which will likely dominate the total uncertainty budget.
In particular, significant R\&D is required to reduce the uncertainty in the calibration of ground-based detectors from the current levels (few percent~\cite{2017PhRvD..96j2001C}) to what is required to maximize the scientific potential of 3G detectors.

For similar reasons, high SNRs will require extremely faithful waveforms, including surrogate models. Techniques used to speed up the likelihood calculations, such as reduced order quadratures, must similarly be tuned to avoid unwanted systematic errors, Appendix~\ref{App.Systematics}.

\section{Conclusions}\label{sec:conclusions}

With several binary black holes  and a binary neutron star detected during Advanced LIGO and Virgo's first two observing runs, the prospects are great for advanced ground-based gravitational-wave detectors: dozens of compact binaries will be detected every year in the local universe.

Future upgrades of the current facilities can increase the strain sensitivity by a factor of \si 2 beyond advanced detectors (Voyager design), whereas the proposed 3G ground-based detectors would provide another order of magnitude increase in sensitivity.
The Einstein Telescope 3G design has been the first one to be proposed. With three 10 km arms forming a triangle of interferometers, it would be able to detect sources at high signal-to-noise ratio, and to yield polarization measurements (which cannot be done by a single L-shaped detector).
The other main design, Cosmic Explorer, keeps LIGO's shape, but with the arms of the interferometer extended to 40 km.
These third-generation detectors would be capable of detecting events to large redshifts ($z\si10$), as well as providing much more accurate parameter estimation of nearby events.

With the construction of 3G instruments years away, a firm timeline does not exist. The order in which the instruments will be built, their location and orientation are still being debated. On the other hand, given their limited cost, the upgrade of existing facilities to Voyager-class seems a no-brainer.

As the first 3G detector comes online, the question arises of whether Voyagers detectors would yield significant useful science on top of what is already provided by a single third-generation observatory, or should instead be decommissioned.

In this paper we attempt to evaluate the usefulness of a heterogeneous network in comparison with both a single third-generation detector and full third-generation networks.
We consider five hypothetical networks, intending to observe the effects of adding LIGO Voyagers to a network that otherwise consists of a single Einstein Telescope.
We generated a population of binary black hole events, with source-frame total masses drawn uniformly from the $[12,200]$ M$_\odot$ range, random spins and a uniform distribution in space. 
For the events with signal-to-noise above the detection threshold, parameter estimation was performed with stochastic sampling, yielding posterior distributions for each parameter.

We found that the sky localization benefited the most with the inclusion of previous-generation detectors, with improvements of many orders of magnitude for relatively nearby events ($z\lesssim3$).
Measurement of the luminosity distance saw smaller improvement with Voyagers, and more significant ones if another third-generation detector is added. The orbital inclination angle, usually unmeasurable with a single ET, can instead be constrained if even a single Voyager is added, at least for nearby events.

The measurement of intrinsic parameters, detector-frame masses and spin is marginally sensitive to the number of detectors in the network, particularly when the extra detectors are Voyagers.
A second third-generation detector is required to see any significant improvement in spin constraints, with a Cosmic Explorer performing slightly better for more distant events.
Source-frame masses can be more precisely estimated adding one or two Voyagers, due to the improvement in luminosity distance, necessary to convert from detector-frame to source-frame mass.

The astrophysical population we generated in the first part of this paper did not result in any source with $z\lesssim 0.3$, since much more volume is available at redshifts of $1-2$. 
To check how well one could characterize CBCs at redshifts of $z\sim 0.1$, similar to those detected by Advanced LIGO and Virgo, we have generated software replicas of GW150914 and GW151226, and analyzed them with the networks of Voyagers and 3Gs.
We found that although Voyager detectors would not provide a large signal-to-noise ratio on the top of what a single ET can deliver, they can improve the estimation of the sky localization by one or two orders of magnitude, depending on whether one or two Voyager sites are added to a single ET.
As for the rest of the BBH population, characterization of intrinsic parameters does not significantly improve.

A GW150914-like source detected by an ET-Voyager network would be localized to an volume typically containing only one galaxy, and with a distance uncertainty of the order of \si $1\%$. Redshift measurement of the single galaxy compatible with the error volume can be used, together with the luminosity distance estimation by gravitational-wave detectors to yield a measurement of the Hubble constant. 
As $\sim100$ sources are detectable  per year at redshifts of $0.1$, an extremely precise measurement of the Hubble constant could be made using these loud events.

\section{Acknowledgments}

The authors acknowledge the support of the National Science Foundation and the LIGO Laboratory. LIGO was constructed by the California Institute of Technology and Massachusetts Institute of Technology with funding from the National Science Foundation and operates under cooperative agreement PHY-0757058.
The authors would like to acknowledge the LIGO Data Grid clusters, without which the simulations could not have been performed. Specifically, we thank the Albert Einstein Institute in Hannover, supported by the Max-Planck-Gesellschaft, for use of the Atlas high-performance computing cluster.

We thank Emanuele Berti, Juan Calderon Bustillo, Hsin-Yu Chen, Evan Hall, Carl-Johan Haster, Scott Hughes, and Colm Talbot for useful discussion
We thank Eve Chase for pointing out an embarrassing typo in the first draft of this paper. 
This is LIGO Document P1800073.

\appendix
\section{ROQ systematics}\label{App.Systematics}

At first, we ran the simulations for the golden events in Sec.~\ref{sec:nearby} using the reduced order quadrature (ROQ) approximation of the likelihood~\cite{Smith:2016qas}.
This is a standard tool, which has been used by the LIGO and Virgo collaborations in all of the  BBH detection papers, and can yield a significant speed-up, especially for long signals.

\begin{figure*}[htb]
\begin{tabular}[c]{cc}
\centering
\begin{subfigure}{0.4\textwidth}
\includegraphics[width=\textwidth]{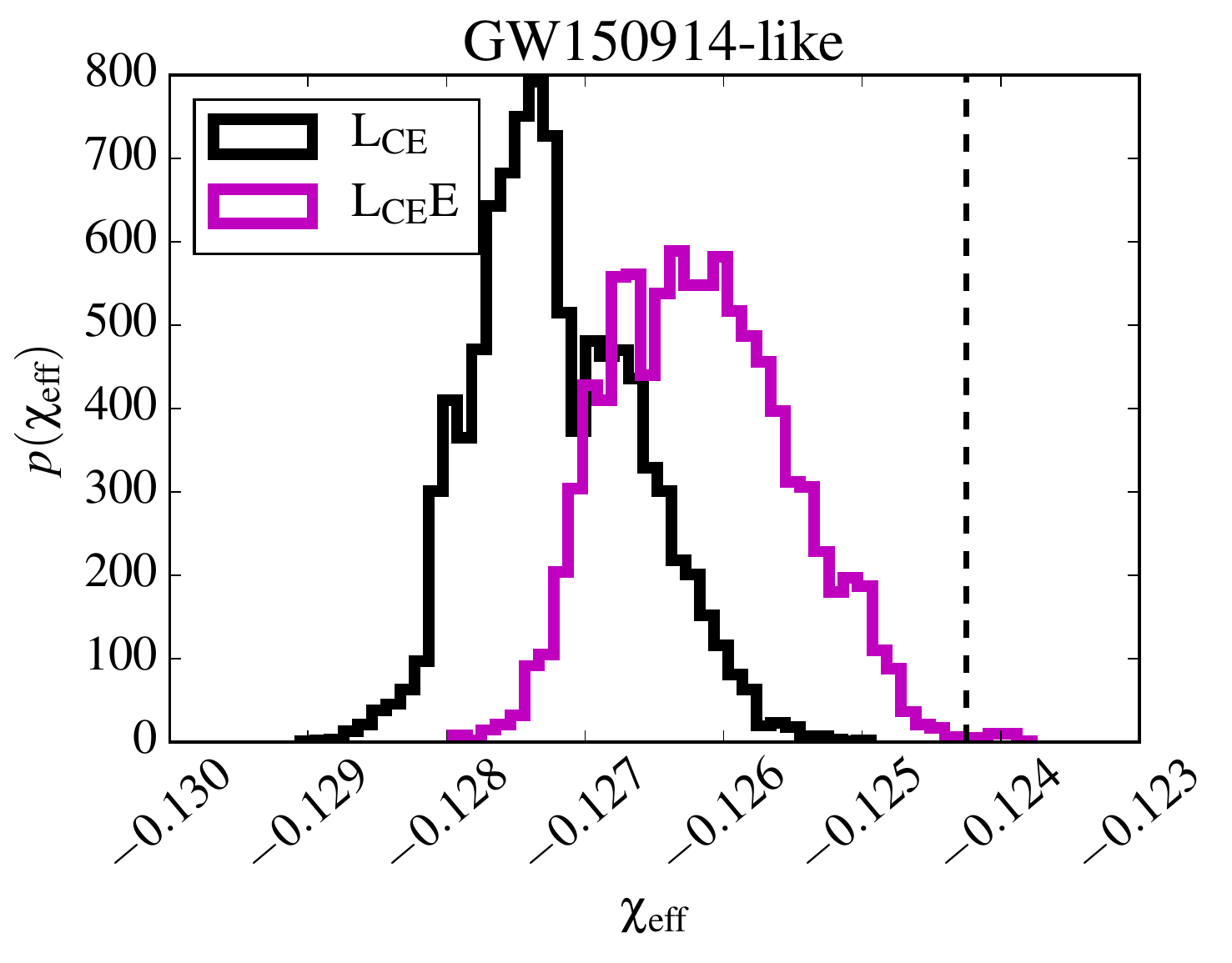}
\end{subfigure} & 
\begin{subfigure}{0.4\textwidth}
\includegraphics[width=\textwidth]{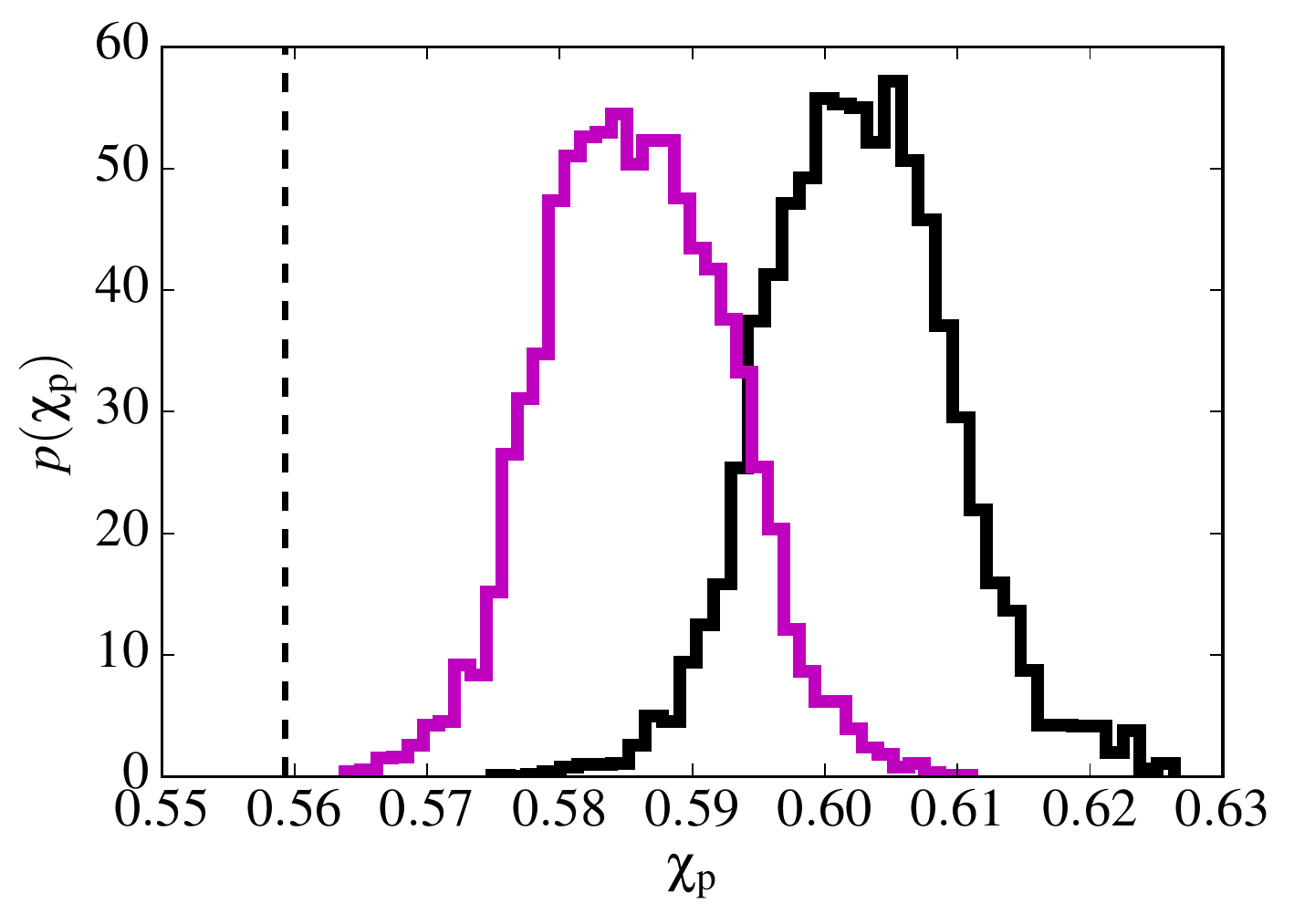}
\end{subfigure} \\
\begin{subfigure}{0.4\textwidth}
\includegraphics[width=\textwidth]{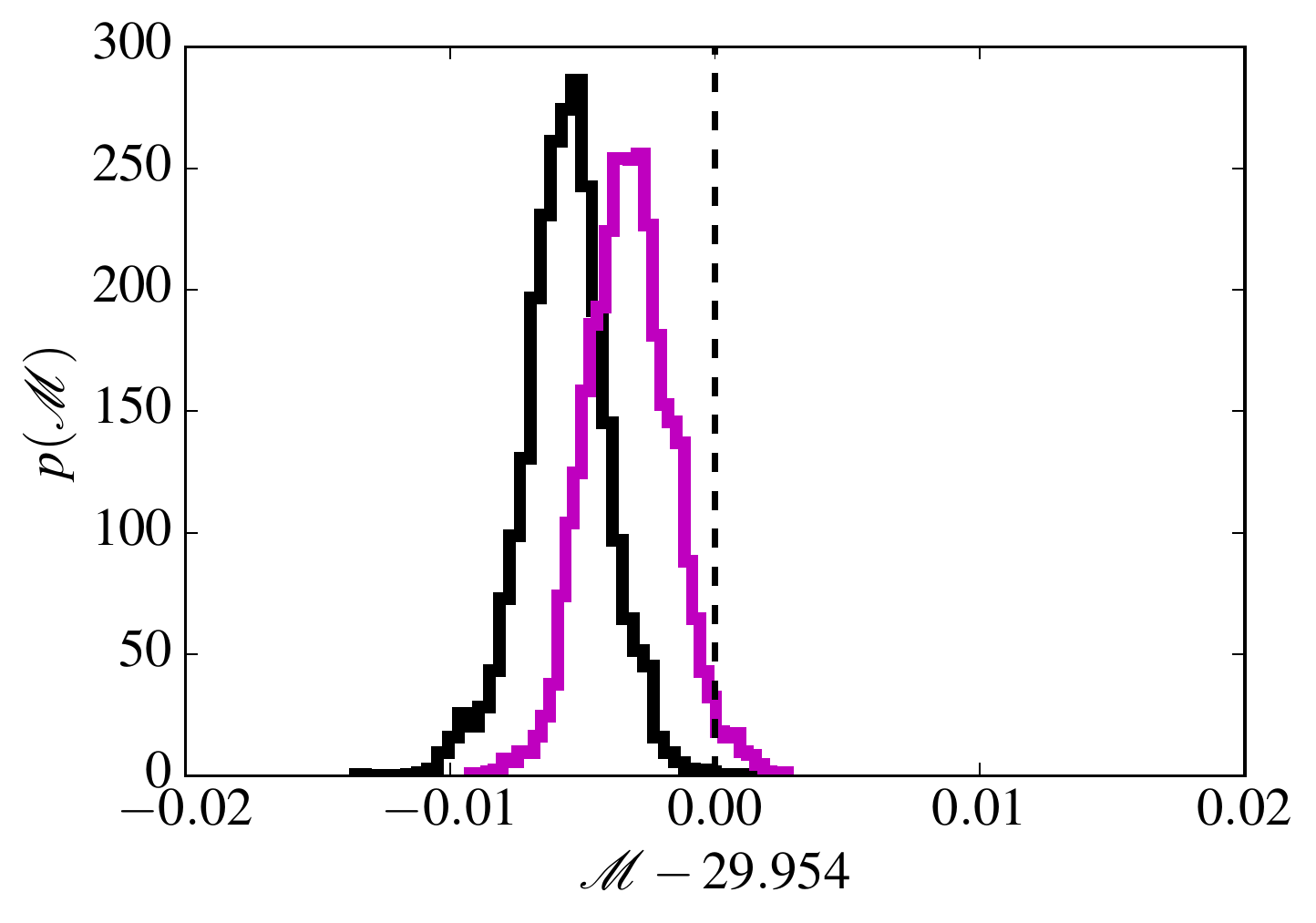}
\end{subfigure} & 
\begin{subfigure}{0.4\textwidth}
\includegraphics[width=\textwidth]{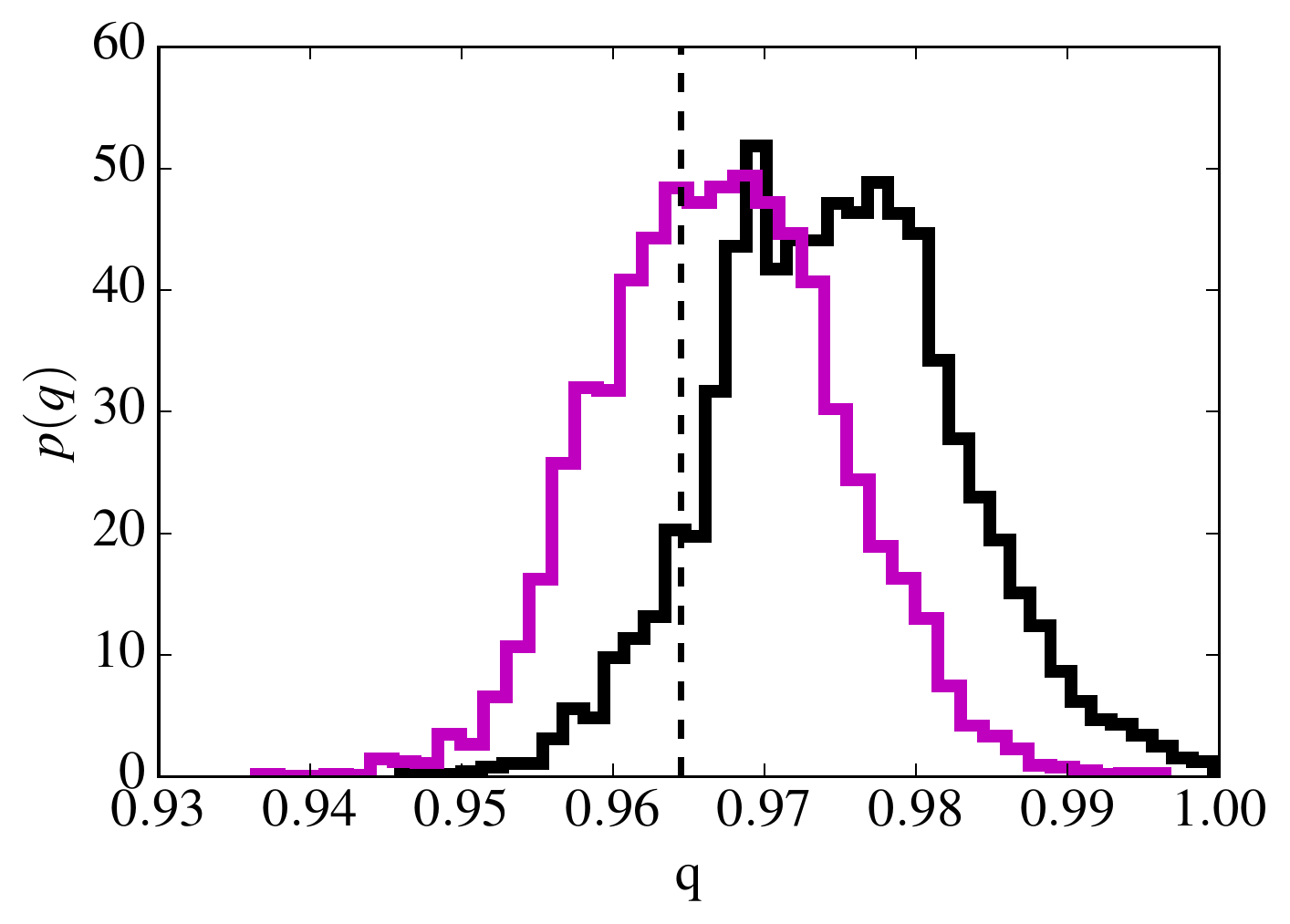}
\end{subfigure} 
\end{tabular}\caption{Posterior distributions for a GW150914-like signal (left column) obtained using the reduced order quadrature approximation of the likelihood with settings adequate for the 2G era. A clear bias if visible}\label{Fig.ReplicaWrong}
\end{figure*}

The posteriors we obtained showed clear biases when the SNR is high enough, Fig.~\ref{Fig.ReplicaWrong}. This is because the approximations used while building the reduced order quadrature basis can introduce systematics errors.

Those are absolutely negligible for events with SNRs of tens - that is, all events detected so far, and the overwhelming majority of sources detectable by 2G detectors - but  can become relevant for the extermely loud events we considered in that section.

The results presented in Sec.~\ref{sec:nearby} thus do not use the reduced order quadrature approximation.

While this issue could have been avoided simply by producing more dense basis for the ROQ, we mention our original attempt to stress how any potential systematic can play a major role in 3G science.
Systematics errors small enough not to produce any visible effect while analyzing signals with advanced detectors, can instead be dominant for 3G instruments.

As the scientific community moves toward identifying the science goals that can be tackled by 3G detectors, waveform development and data analysis techniques  must follow.

\bibliography{SalvosBib}
\end{document}